\documentclass[review]{elsarticle}

\usepackage{lineno,hyperref}
\usepackage{mathtools}
\usepackage{amsmath}
\usepackage{color}
\usepackage{xcolor}
\usepackage{bm}
\usepackage{mathrsfs}
\usepackage{amssymb}
\usepackage{euscript}
\usepackage{algorithmic}
\usepackage{graphicx}
\usepackage{subfigure}
\usepackage{multirow}
\usepackage[Symbol]{upgreek}
\usepackage{booktabs}
\usepackage{makecell}
\usepackage[english]{babel}
\usepackage{diagbox}
\usepackage{bbding}
\usepackage{cancel}
\usepackage{rotating}

\modulolinenumbers[1]

\journal{Speech Communication}

\begin{document}

\begin{frontmatter}

\title{A Two-stage Complex Network using Cycle-consistent Generative Adversarial Networks for Speech Enhancement}

\address[mymainaddress]{State Key Laboratory of Media Convergence and Communication, Communication University of China, 100024, Beijing, China}
\address[mysecondaryaddress]{Key Laboratory of Media Audio and Video, Ministry of Education, Communication University of China, 100024, Beijing, China}
\address[mythirdaddress]{Key Laboratory of Noise and Vibration Research, Institute of Acoustics, Chinese Academy of Sciences, 100190, Beijing, China}
\address[myfourthaddress]{University of Chinese Academy of Sciences, 100049, Beijing, China}
\cortext[mycorrespondingauthor]{Corresponding author: Hui Wang (hwang@cuc.edu.cn) and Chengshi Zheng (cszheng@mail.ioa.ac.cn)}

\author[mymainaddress,mysecondaryaddress, mythirdaddress]{Guochen Yu}
\author[mymainaddress,mysecondaryaddress]{Yutian Wang}
\author[mymainaddress,mysecondaryaddress]{Hui Wang}
\author[mymainaddress,mysecondaryaddress]{Qin Zhang}
\author[mythirdaddress,myfourthaddress]{Chengshi Zheng}

\begin{abstract}
Cycle-consistent generative adversarial networks (CycleGAN) have shown their promising performance for speech enhancement (SE), while one intractable shortcoming of these CycleGAN-based SE systems is that the noise components propagate throughout the cycle and cannot be completely eliminated. Additionally, conventional CycleGAN-based SE systems only estimate the spectral magnitude, while the phase is unaltered. Motivated by the multi-stage learning concept, we propose a novel two-stage denoising system that combines a CycleGAN-based magnitude enhancing network and a subsequent complex spectral refining network in this paper. Specifically, in the first stage, a CycleGAN-based model is responsible for only estimating magnitude, which is subsequently coupled with the original noisy phase to obtain a coarsely enhanced complex spectrum. After that, the second stage is applied to further suppress the residual noise components and estimate the clean phase by a complex spectral mapping network, which is a pure complex-valued network composed of complex 2D convolution/deconvolution and complex temporal-frequency attention blocks. Experimental results on two public datasets demonstrate that the proposed approach consistently surpasses previous one-stage CycleGANs and other state-of-the-art SE systems in terms of various evaluation metrics, especially in background noise suppression.

\end{abstract}


\end{frontmatter}


\section{Introduction}
Speech enhancement (SE) can be described as the technique to separate the speech components from the background noise interference. It intends to improve speech quality and intelligibility in many communication applications such as front-ends for automatic speech recognition (ASR) systems and hearing aids~{\cite{loizou2013speech}}. In recent years, due to the tremendous ability of deep neural networks (DNNs) to deal with non-stationary noise in low signal-to-noise ratio (SNR) conditions, many DNN-based approaches have demonstrated superior performance in single-channel SE~{\cite{wang2018supervised}}. These DNN-based methods can be divided into two categories, namely masking-based approaches~{\cite{wang2014training, hummersone2014ideal}} and mapping-based approaches~{\cite{lu2013speech, xu2014regression, yuan2020time, meng2018adversarial, liao2018noise}}. The masking-based approaches estimate a time-frequency (T-F) mask that is applied to a noisy speech signal for enhancement (e.g., ideal binary mask (IBM), ideal ratio mask (IRM)). The mapping-based approaches are proposed to train a mapping network that directly transforms noisy speech features (e.g., magnitude spectra, log-power spectra) to clean ones.

More recently, generative adversarial networks (GANs)~{\cite{goodfellow2014generative}} have demonstrated comparable performance for SE. The same as most other DNN-based approaches, they also require a large number of paired training data, which may be difficult for practical applications. To solve the difficulty of obtaining the parallel recordings of speech and noise from the real scenarios, it is suggested using cycle-consistent GAN (CycleGAN)~{\cite{zhu2017unpaired}} for SE. Moreover, CycleGAN-based approaches have also demonstrated their promising performance with parallel data, for their capability of preserving the speech structure and reducing speech distortion~{\cite{meng2018cycle,xiang2020parallel, wang2020improved}}.

Nevertheless, conventional CycleGAN-based approaches have two intractable limitations for SE tasks. Firstly, to ensure cycle-consistency of the original noisy speech domain and target clean speech domain, the enhanced signal always contains the original noise information. In other words, the cycle-consistency-based methods remain audible residual noise, which is challenging to eliminate. This essentially implies that background noise reduction is challenging in these algorithms. Secondly, previous CycleGAN-based SE systems only estimated the magnitude spectrum, log-power spectrum or Mel-cepstral coefficients features while combining the non-oracle (i.e., noisy) phase to reconstruct the time-domain waveform. This leads to severe phase distortion under low SNR scenarios, resulting in serious performance degradation in speech quality and intelligibility.

Multi-stage learning approaches can decompose the original difficult task into multiple more manageable sub-tasks and have demonstrated better performance than single-stage methods in many areas, such as image inpainting~{\cite{hedjazi2020texture}} and image deraining~{\cite{li2018recurrent}}. In this paper, we incorporate a CycleGAN-based magnitude spectrogram mapping network (dubbed CycleGAN-MM) and a deep complex-valued denoising network (dubbed DCD-Net) as a two-stage approach for SE. In the first stage, we utilize CycleGAN-MM to only estimate the magnitude of clean spectra, where we introduce a relativistic average loss in discriminators to stabilize the training. Motivated by recent studies in SE area for better sequence modeling~{\cite{tang2020joint, zheng2020interactive}}, we employ temporal-frequency attention (T-FA) in generators to capture global dependency along temporal and frequency dimensions, respectively. More recently, it has been demonstrated that the magnitude and phase are difficult to be optimized simultaneously, especially in extremely low SNR conditions~{\cite{wang2020complex}}. In our preliminarily investigated one-stage CycleGAN-based complex mapping SE system (dubbed CycleGAN-CM), optimizing both real and imaginary (RI) components may cause the unstable training of the generators and discriminators, and consequently its performance get even worse than only estimating the magnitude of clean speech spectrum. That is because when optimizing phase information by estimating the complex spectrum, magnitude estimation may deviate its optimal convergence path by degrees~{\cite{li2021icassp}}. Therefore, we only optimize the magnitude of the clean spectra in the first stage, which is then coupled with its corresponding noisy phase to obtain a coarsely enhanced complex spectrum.
Subsequently, in the second stage, we introduce a deep complex denoising net to further suppress the intractable remaining noise in the previous stage, while simultaneously reconstructing the clean speech phase by estimating both real and imaginary components of the clean spectra. Instead of using real-valued neural networks, DCD-Net is designed by complex-valued convolutional networks and complex T-FA blocks to refine the coarsely enhanced complex spectrum. To the best of our knowledge, this is the first attempt to handle the complex spectral mapping in CycleGAN-based SE systems. To validate the superiority of the proposed scheme, we compare our model with recent state-of-the-art (SOTA) GAN-based and Non-GAN-base SE systems on two public datasets. Experimental results demonstrate the proposed two-stage approach outperforms the one-stage CycleGAN-based systems by a significant margin and achieves SOTA performance.

The remainder of this paper is organized as follows. Section~{\ref{Sec2}} introduces the related works, including CycleGAN for SE, deep complex-valued SE systems, and multi-stage SE approaches. In Section~{\ref{Sec3}}, the proposed framework is described in detail. The experimental setup is presented in Section~{\ref{Sec4}}, and the experimental results and analysis are provided in Section~{\ref{Sec5}}. Finally, some conclusions are drawn in Section~{\ref{Sec6}}.

\section{Related works\label{Section2}}
\label{Sec2}
\subsection{Cycle-consistent GAN-based for SE\label{Section21}}
A recent breakthrough in the SE area comes from the application of GANs as feature mapping networks. GAN consists of a generator network ($G$) and a discriminator network ($D$) that play a min-max game between each other. By using adversarial training, the objective of $G$ is to synthesize the fake samples which are indistinguishable from the target data distribution, whilst $D$ attempts to discriminate between the real and fake samples.
SEGAN is the pioneering work employing GAN for SE task, which directly maps raw waveform of the clean speech from the mixed raw waveform in time domain~{\cite{pascual2017segan}}. More recently, Other GAN-based SE algorithms in the time domain have been proposed to leverage different loss functions~{\cite{baby2019sergan, fu2019metricgan}} or generator structures~{\cite{liu2020cp, pascual2019time}}. Another mainstream of GAN-based SE algorithms operates on the time-frequency (T-F) domain, where $G$ is used to estimate a T-F mask~{\cite{soni2018time}}.

\begin{figure}[t]
	\centering
	\centerline{\includegraphics[width=80mm]{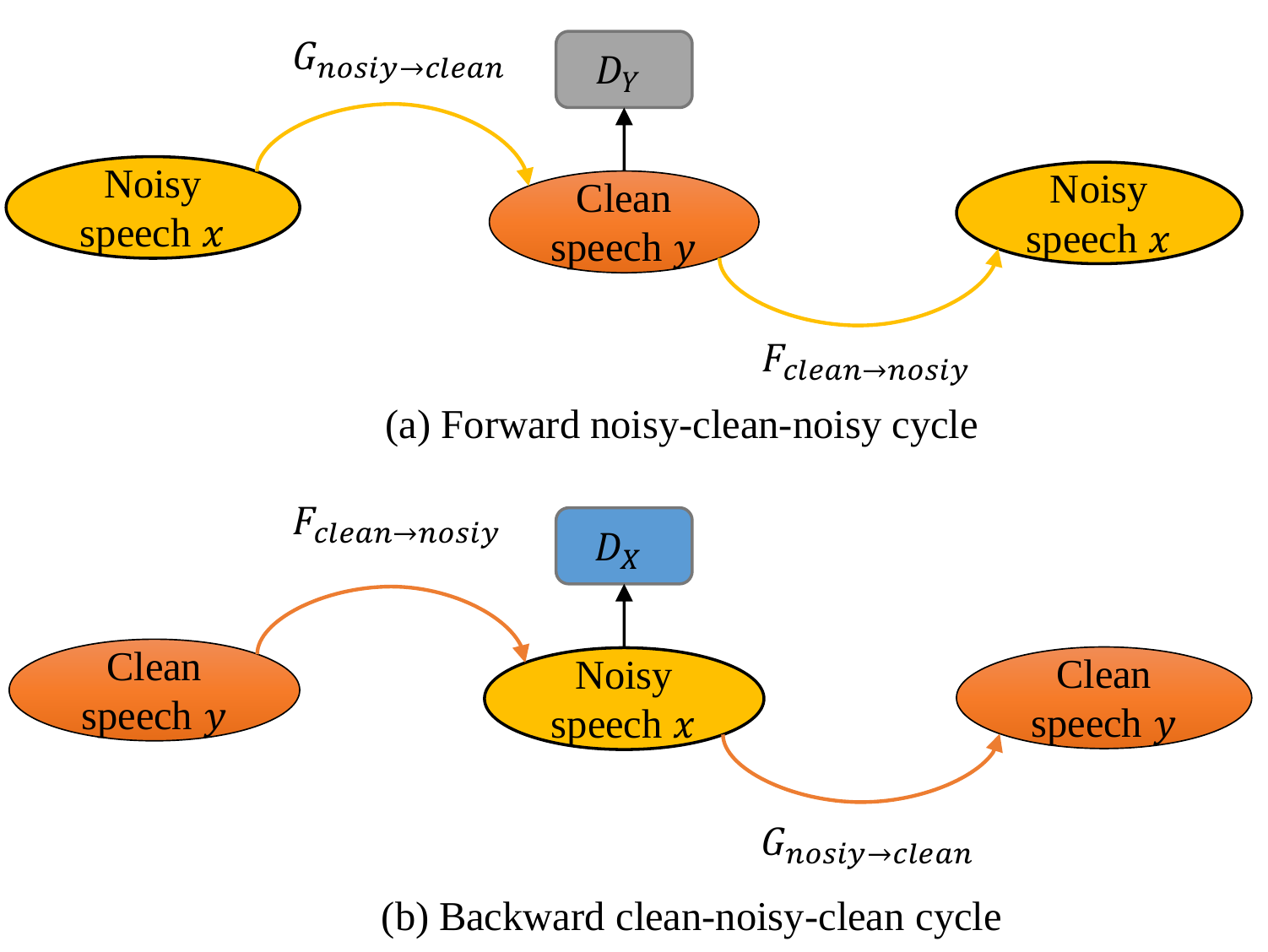}}
	\caption{(a) The forward noisy-clean-noisy cycle. (b) The backward clean-noisy-clean cycle.}
	\label{fig:cyclegan}
	\vspace{-0.2cm}
\end{figure}

However, conventional GAN-based methods may map noisy features to any random permutation of the clean features in the target domain with only adversarial losses, thus there is no guarantee that the individual enhanced feature is exactly paired with the target clean one~{\cite{zhu2017unpaired}}. In other words, these methods cannot restrict that the contextual information of noisy features and enhanced features are always cycle-consistent. As a variant of GAN, CycleGAN is widely used for the unpaired image-to-image translation task, while in speech area it also demonstrates excellent performance on voice conversion~{\cite{kaneko2018cyclegan}}, music style transfer~{\cite{brunner2018symbolic}}, and SE~{\cite{meng2018cycle,xiang2020parallel, wang2020improved}}. Incorporating CycleGANs for SE, these methods demonstrate their effectiveness in improving SE performance especially in maintaining speech integrity and reducing speech distortion. CycleGAN-based approaches for SE contain a noisy-to-clean generator $G$ and an inverse clean-to-noisy generator $F$, which transforms the noisy features into the enhanced ones for the former, and vice versa for the latter. As illustrated in Fig.~{\ref{fig:cyclegan}}, a forward noisy-clean-noisy cycle and a backward clean-noisy-clean cycle jointly constrain $G$ and $F$ to be cycle-consistent, which are optimized with the adversarial loss, a cycle-consistency loss, and an identity-mapping loss, respectively. Discriminators $D_X$ and $D_Y$ are trained to classify the target speech features as real and the generated speech features as fake.

Nevertheless, in the standard CycleGAN-based approaches, the enhanced signal always contains the original noise information and remains audible residual background noise due to the constraint of cycle-consistency. Moreover, to the best of our knowledge, phase recovery has not been well investigated in previous CycleGAN-based SE approaches.

\subsection{Two-stage approaches for noise reduction\label{Section22}}
Recently, some studies focus on conducting the original SE task by multi-stage networks, which can significantly improve the estimated speech quality. In~{\cite{hao2020masking}}, Hao et al. proposed a masking and inpainting SE approach for low SNR and non-stationary noise. This two-stage approach consists of binary masking and spectrogram inpainting. In the first stage, a binary masking model is trained to remove the T-F points dominated by severe noise in low SNR conditions and obtain the spectra with the T-F points dominated by clean speech, while an inpainting model is used to recover the missing T-F points in the second stage. To solve the difficulty of estimating phase spectrum, Du et al. introduced a joint framework composed of a Mel-domain denoising autoencoder and a deep generative vocoder for monaural speech enhancement\cite{du2020joint}, in which the clean speech waveform is reconstructed without using the phase. The first stage enhances the Mel-power spectrum of noisy speech by denoising autoencoder, while a deep generative vocoder focuses on synthesizing the speech waveform in the second stage. More recently, Li et al. proposed a two-stage approach named CTS-Net for SE in complex domain~{\cite{li2021icassp}}, in which a coarse magnitude estimation network (CME-Net) and a complex spectrum refined network (CSR-Net) jointly optimize the noisy spectra. In the first stage, the target spectral magnitude is coarsely estimated, which is then coupled with the noisy phase to obtain a coarsely estimated complex spectrum. In the second stage, CSR-Net is trained to estimate both RI components of the clean complex spectrum, thus further reducing the residual noise, restoring the clean speech phase, and inpainting missing details of the estimated spectrum.

\subsection{Deep complex neural network based for phase-aware SE\label{Section23}}

Conventional SE methods only estimate the magnitude of the speech, and the time-domain speech waveform is reconstructed by using the noisy phase and the estimated magnitude. The reason for not enhancing the phase is that it was believed that phase is not so important for SE~{\cite{wang1982unimportance}}, as well as it is intractable to directly estimate the clean speech phase because it is unstructured. However, recent studies~{\cite{paliwal2011importance}} show the importance of the accurate phase as it can significantly improve perceptual speech quality, especially in low SNR conditions. Subsequently, some SE algorithms ~{\cite{mowlaee2012phase, kulmer2014phase}} are developed to solve this problem, and consistently show objective speech quality improvements when the phase is enhanced.

More recently, the deep learning-based phase-aware SE algorithms can be divided into two categories. The first one operates in the time domain by estimating speech signals from raw-waveform noisy signals without using any explicit T-F representation~{\cite{pascual2017segan, baby2019sergan, pascual2019time, pandey2020densely}}, thereby avoiding the problem of phase estimation. Recently, it is reported that the fine-detailed structures of noise and speech components are more separable with T-F representation. Hence, another main-stream phase-aware SE approaches work on optimizing both RI components of the complex spectrum by using complex-valued ratio mask (CRM) ~{\cite{williamson2015complex}} or a direct complex-mapping network~{\cite{tan2019complex, tan2019learning, zheng2020interactive}}. Thereby, these methods estimate both magnitude and phase information in the frequency domain. Although the above approaches have been proposed to address this issue, they are limited as the neural network still conducts real-valued operations. To this end, deep complex u-net (named DCUNET)~{\cite{choi2019phase}} and Deep Complex Convolution Recurrent Network (named DCCRN)~{\cite{hu2020dccrn}} are proposed to conduct phase-aware SE via complex-valued neural network~{\cite{trabelsi2017deep}}. DCUNET incorporates a deep complex network and u-net structure to enhance the complex spectra of noisy speech. Note that DCUNET is trained to estimate bounded CRM and optimizes the weighted source-to-distortion ratio (wSDR) loss~{\cite{venkataramani2017adaptive}} after reconstructing the enhanced time-domain waveform by inverse STFT. DCCRN effectively combines both the advantages of DCUNET and CRN~{\cite{tan2018convolutional}} to estimate the complex spectra of clean speech, in which LSTM is utilized to model temporal context with significantly reduced trainable parameters. However, it may cause the memory bottleneck issue when using LSTM to model temporal dependencies, which may result in reducing training efficiency.

Motivated by these studies, we propose a two-stage deep complex approach, which incorporates a complex-valued denoising network (named DCD-Net) with a CycleGAN-based magnitude mapping network (named CycleGAN-MM). Specifically, CycleGAN-MM is adopted to coarsely estimate the clean spectral magnitude in the first step, while DCD-Net aims to further suppress the intractable residual background noise and simultaneously recover the clean phase information implicitly by estimating both RI components of the clean spectrum.

\begin{figure*}[t]
	\centering
	\centerline{\includegraphics[width=120mm]{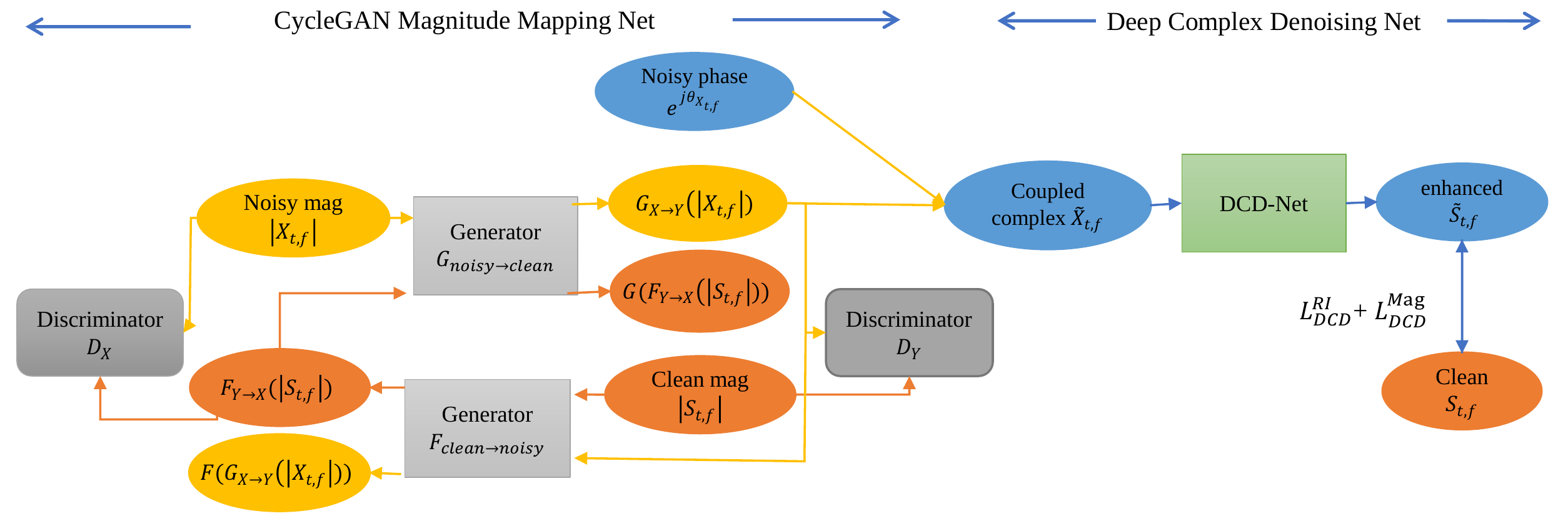}}
	\caption{
		The diagram of the proposed CycleGAN-DCD. The left part denotes the architecture of CycleGAN-MM, while the right part denotes the architecture of DCD-Net.	
	}
	\label{fig:architecture}
\end{figure*}

\section{Method\label{Section3}}
\label{Sec3}
\subsection{Training Target}
The overall network architecture is presented in Fig.~{\ref{fig:architecture}}, which is comprised of two sub-networks, namely CycleGAN-MM and DCD-Net. In our SE task, the mixture signal in the time domain is formulated as ${x(n)}=s(n)+z(n)$, where $x$, $s$ and $z$ denote noisy speech, clean speech and noise, respectively. With the STFT, the noisy speech in the time-frequency domain can be modeled as,
\begin{equation}
	{X_{t,f}}=S_{t,f}+Z_{t,f},
\end{equation}
where $X_{t,f}=\left | X_{t,f} \right |e^{j\theta_{X_{t,f}}}\in \mathbb{C}$, $S_{t,f}=\left | S_{t,f} \right |e^{j\theta_{S_{t,f} }}\in \mathbb{C}$ and $Z_{t,f}=\left | Z_{t,f} \right |e^{j\theta_{Z_{t,f} }}\in \mathbb{C}$ denote the time-frequency ($t,f$) representations of noisy speech, clean speech and noise, respectively. The input to CycleGAN-MM is the magnitude of the noisy spectrum $\left | \widetilde{X}_{t,f} \right |=G(\left | {X}_{t,f} \right |)$. After the first-stage, the estimated magnitude $\left | \widetilde{X}_{t,f} \right |$ is then coupled with the original noisy phase $e^{j\theta_{ X_{t,f} }}$ to obtain a coarsely enhanced complex spectrum $\widetilde{X}_{t,f}$. In the second stage, DCD-Net receives the enhanced complex spectrum to estimate the CRM, which can be defined as:
\begin{equation}
	{CRM}=\frac{X_rS_r+X_iS_i}{X^{2}_r+X^{2}_i}+j\frac{X_rS_i-X_iS_r}{X^{2}_r+X^{2}_i}= \widetilde{M}_{r} +j \widetilde{M}_{i}
\end{equation}
where $X_r$ and $X_i$ denote the RI components of the noisy speech spectrum, respectively. Similarly, $S_r$ and $S_i$ indicate the RI components of the clean speech spectrum. The real and imaginary parts of the CRM are represented by $\widetilde{M}_{r}$ and $\widetilde{M}_{i}$. Alternatively, the polar coordinate representation of $\widetilde{M}$ can be presented as,
\begin{equation}	
	{{\widetilde{M}}}=\widetilde{M}_{mag}\cdot e^{j\theta_{\widetilde{M}_{phase}}}
\end{equation}
where $\widetilde{M}_{mag}$ and $\theta_{\widetilde{M}_{phase}}$ denote the magnitude and phase of the complex-valued mask, respectively. To make the mask bounded in an unit-circle at the complex space, we use the $tanh$ activation function to limit the magnitude mask $\widetilde{M}_{mag}$ ranging from 0 to 1 like in~{\cite{choi2019phase}}. Hence, the final estimated clean complex spectrum $\widetilde{S}$ of DCD-Net in polar coordinates can be calculated by,

\begin{equation}
	{{\widetilde{S}}}=\left | \widetilde{X}_{t,f} \right | \cdot\widetilde{M}_{mag}\cdot e^{j(\theta _{\widetilde{X}_{t,f}} +\theta _{\widetilde{M}_{phase}})}.
\end{equation}

Note that DCD-Net is optimized by signal approximation (SA), which directly minimizes the difference between the complex spectrum of the clean speech and that of the noisy speech applied with bounded CRM.

\begin{figure*}[t]
	\centering
	\centerline{\includegraphics[width=120mm]{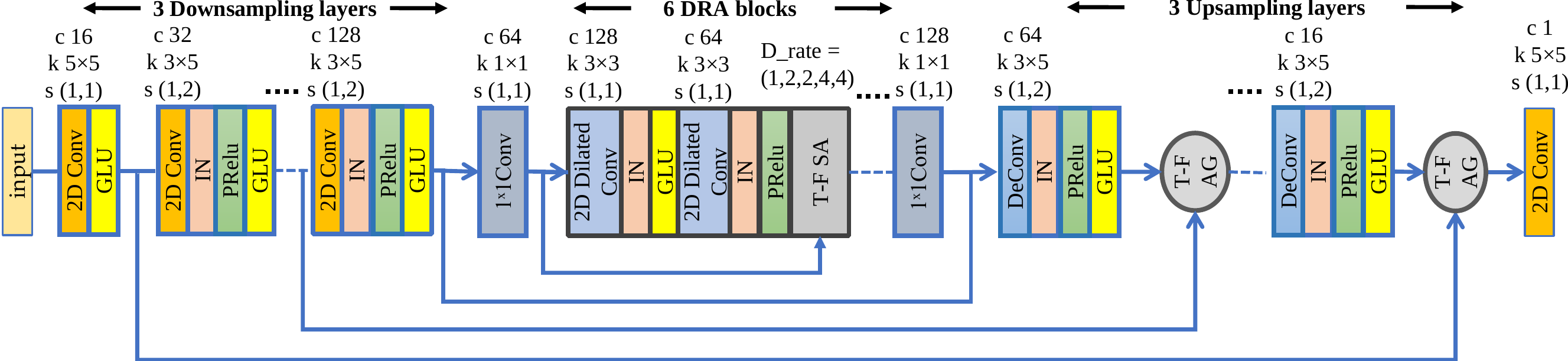}}
	\caption{
		The generator of CycleGAN-MM.
		$c$, $s$ and $k$ represent output channel numbers, kernel size, and stride of 2D convolution, respectively. IN, PRelu, GLU indicate instance normalization, gated linear unit, and parametric Relu activation, respectively.
	}
	\label{fig:generator}
\end{figure*}

\subsection{CycleGAN-based Magnitude Mapping network}

As shown in Fig.~{\ref{fig:architecture}}, two generators, dubbed $G_{X\rightarrow Y}$ and $F_{Y\rightarrow X}$, and two discriminators, dubbed $D_X$ and $D_Y$, are employed in CycleGAN-MM simultaneously. The generator is composed of three components, namely three downsampling layers, six dilated residual attention (DRA) blocks and three homologous upsampling layers, where the detailed structure of this generator is illustrated in Fig.~{\ref{fig:generator}}. Each downsampling/upsampling layer block is composed of a 2D convolution/deconvolution layer, followed by instance normalization (IN), parametric Relu activation function (PRelu) and gated liner units (GLUs). GLUs can control the information flows throughout the network~{\cite{dauphin2017language}}, showing the effectiveness of modeling speech sequential structure. In generators, we introduce Temporal-Frequency self-attention (T-F SA) and Temporal-Frequency attention gates (dubbed T-F AG) in DRA block and upsampling layers, respectively. T-F SA and T-F AG are utilized herein to capture relative contextual dependencies along the time and frequency dimensions and directly pass the salient information of source speech features. The discriminator is composed of six 2D convolutions, each of them followed by spectral normalization (SN) and PRelu, compressing the feature maps into a high-level representation. SN is proposed to stabilize the training process of the discriminator~{\cite{miyato2018spectral}} and demonstrates the effectiveness to avoid vanishing or exploding gradients. The kernel size for each 2D convolutions layer is (3, 5) except (1, 1) for the last layer in the temporal and frequency axis, respectively. The stride is set to (1, 2) along the temporal and frequency axis for the first five downsampling layers, while it is set to (1, 1) for the last layer. The number of channels throughout the 2D convolutions is (32, 32, 64, 64, 128, 1).

\subsubsection{Temporal-frequency attention\label{Section31}}
\label{sec:Section31}
\begin{figure}[t]
	\centering
	\centerline{\includegraphics[width=0.5\columnwidth]{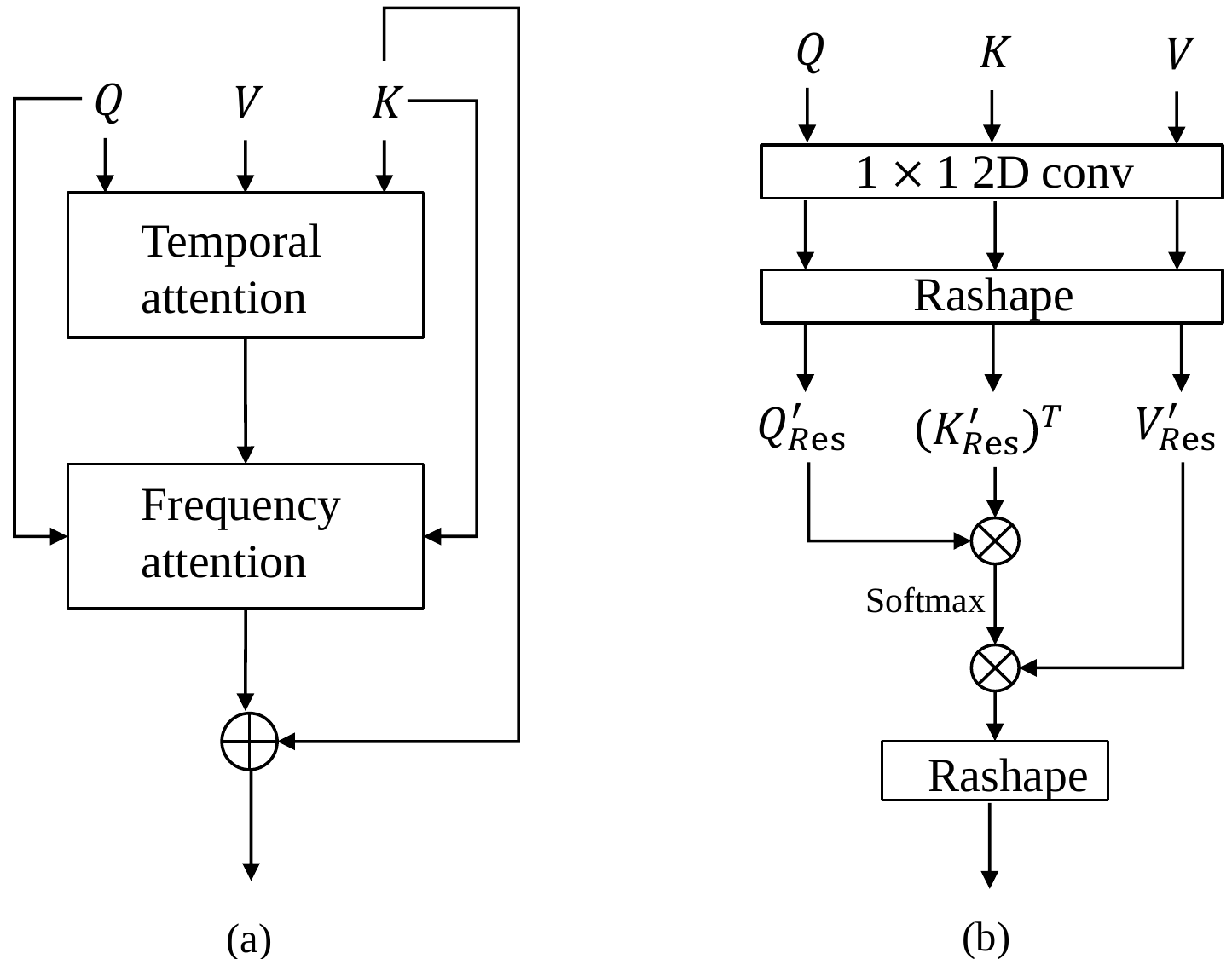}}
	\caption{(a) The diagram of T-FA. (b) The detailed diagram of TA and FA.}
	\label{fig:T-Fatt}
	\vspace{-0.2cm}
\end{figure}

Attention mechanism~{\cite{vaswani2017attention}} has been widely used in speech processing tasks, as it can leverage the contextual information in the time-frequency dimension and further enhance the salient speech information importing in the feature learning procedure. Moreover, using attention in SE task can simulate the human auditory perception, so that more attention is put to speech while less is put to the surrounding background noise~{\cite{anderson2013dynamic}}.
Following the terminology in~{\cite{tang2020joint}}, we compute the attention function on a set of queries, keys, and values simultaneously, and pack them together into feature maps $Q\in \mathbb{R}^{B\times T\times F^{'}\times C}$,$K\in \mathbb{R}^{B\times T\times F'\times C}$ and $V\in \mathbb{R}^{B\times T\times F^{'}\times C}$, respectively. Here, $B$ denotes the batch size of input features, $T$ denotes the number of frames, $F^{'}$ denotes the number of frequency bins and $C$ denotes the number of channels in each feature map. Inside the attention module, the feature maps $Q$ and $K$ are first projected into two feature spaces $Q^{'}$ and $K^{'}$ by $1\times1$ convolutions to calculate attention maps $\beta $. However, the size of the attention weight matrix $\beta $ in the original attention mechanism is $(T\times F^{'})\times (T\times F^{'})$ , which would be extremely large and cost heavy computational complexity. To address this problem, we introduce temporal-frequency attention (T-FA) to capture the global dependencies along temporal and frequency dimensions, respectively. As discussed in~{\cite{tang2020joint, zheng2020interactive}}, by factorizing the original attention into temporal attention (TA) and frequency attention (FA), the large attention weight matrix can be subdivided into two much smaller ones, i.e., $(T\times T)$ and $(F^{'}\times F^{'})$. As shown in Fig.~{\ref{fig:T-Fatt}}, the output $y^t$ of $TA(Q, K, V)$ can be computed as,
\begin{equation}
	\begin{gathered}
		Q^t=W_Q^t\ast Q, K^t=W_K^t\ast K, V^t=W_V^t\ast V,\\
		Q^t_{Res},K^t_{Res},V^t_{Res} = Reshape^t(Q^t,K^t,V^t),\\
		\beta^{t}= softmax((Q^t_{Res})\cdot( K^t_{Res})^{T}),\\
		O^{t} = Reshape^{t'}(\beta^{t}\cdot V^t_{Res}),\\
		y^t = \lambda O^{t}+K,
	\end{gathered}
\end{equation}
where $Q^t_{Res},K^t_{Res},V^t_{Res}\in R^{(B\times F')\times T\times C^t}$, $C^t=\left \{  {\frac{C}{8},\frac{C}{8},C} \right \}$, and $O^{t}\in \mathbb{R}^{B\times T\times F'\times C}$, respectively. Here, $\lambda$ is a learnable scalar coefficient and initialized as 0.
Similarly, $FA(Q, K, V)$ is employed after TA block, which can be expressed as,
\begin{equation}
	\begin{gathered}		
		Q^f=W_Q^f\ast Q,K^f=W_K^f\ast K,V^f=W_V^f\ast V, \\
		Q^f_{Res},K^f_{Res},V^f_{Res} = Reshape^f(Q^f,K^f,V^f),\\
		\beta^{f}= softmax((Q^f_{Res})\cdot( K^f_{Res})^{T}),\\
		O^{f} = Reshape^{f'}(\beta^{f}\cdot V^f_{Res}),\\
		y^f = \lambda O^{f}+K,
	\end{gathered}
\end{equation}
where $Q^f_{Res},K^f_{Res},V^f_{Res}\in R^{(B\times T)\times F'\times C^f}$, $C^f=\left \{  {\frac{C}{8},\frac{C}{8},C} \right \}$, and $O^{f}\in \mathbb{R}^{B\times T\times F'\times C}$, respectively. In the T-F AGs, the memory keys $K$ come from the output of the previous layer or the final DRA block, while the queries $Q$ and values $V$ come from the output of the homologous downsampling layers. For the T-F SA in DRA blocks, all the $Q$, $K$, $V$ come from the same output of the previous dilated residual layers.

\vspace{0cm}

\begin{figure*}[t]
	\centering
	\centerline{\includegraphics[width=120mm]{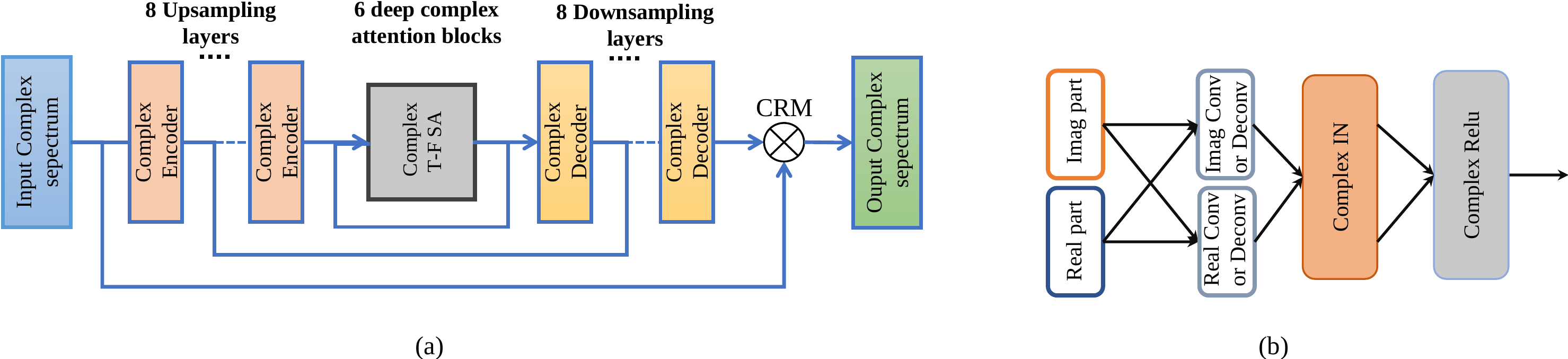}}
	\caption{
		The diagram of deep complex denoising net.
		(a) The topology of DCD-Net. (b) The diagram of complex-valued encoder/decoder.
	}
	\label{fig:dcd-net}
\end{figure*}

\begin{figure}[t]
	\centering
	\centerline{\includegraphics[width=0.9\columnwidth]{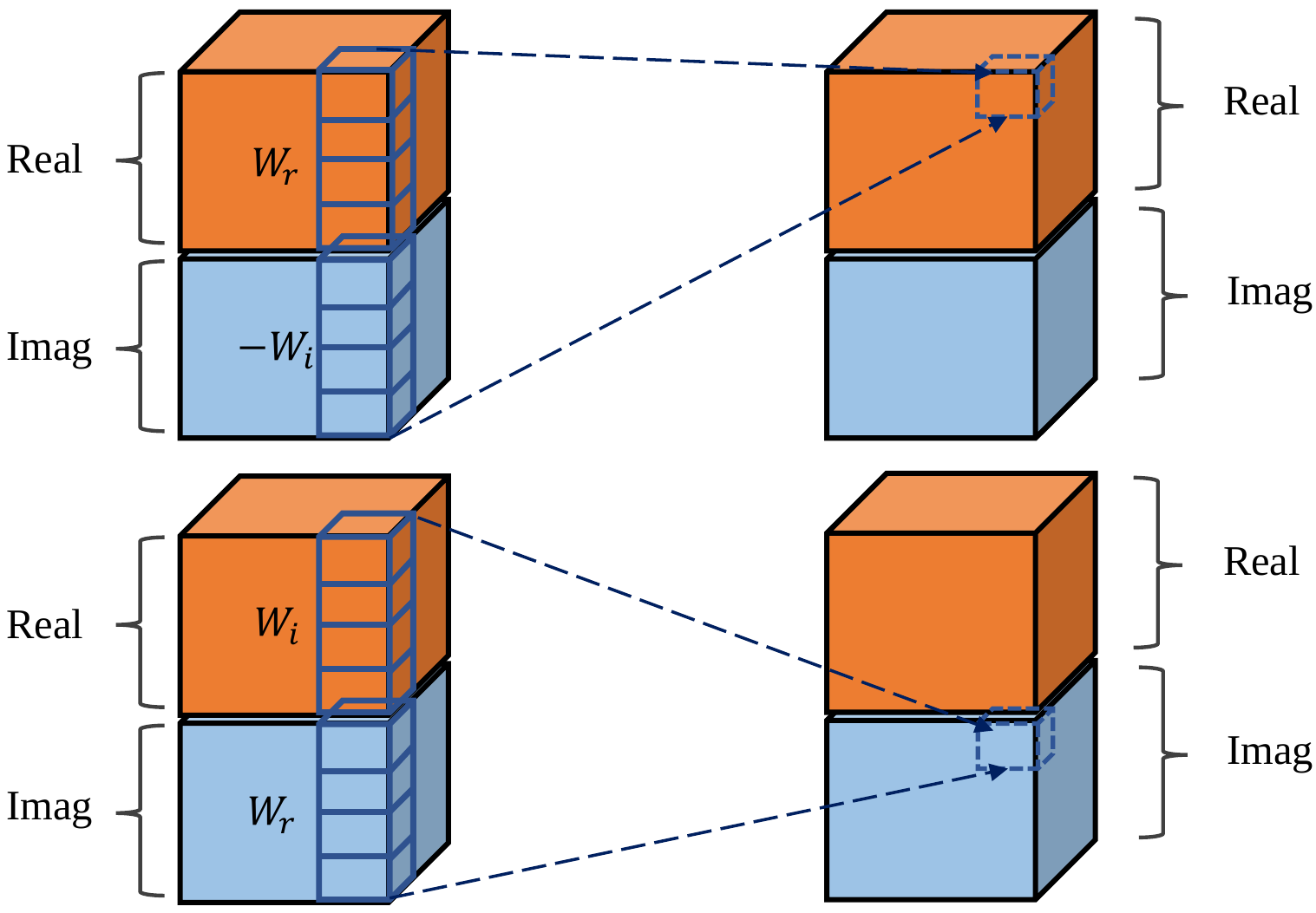}}
	\caption{The diagram of complex-valued convolution network}
	\label{fig:complexcnn}
	\vspace{-0.4cm}
\end{figure}
\vspace{-0.2cm}

\subsubsection{CycleGAN-MM Loss function\label{Section32}}

CycleGAN-MM uses the following three losses to jointly optimize the magnitude estimation process, namely relativistic adversarial losses, cycle-consistency losses, and an identity mapping loss.

\textbf{Relativistic adversarial loss:} For the noisy-to-clean mapping, the relativistic average least-square (RaLS) adversarial loss~{\cite{jolicoeur2018relativistic}} is used to make the enhanced magnitude spectra $G_{X\rightarrow Y}(\left | {X}_{t,f} \right |)$ indistinguishable from the clean ones $\left | {S}_{t,f} \right |$, which can be expressed as below.
\begin{equation}
	\begin{gathered}	
		\mathcal{L}_{Radv}(D_Y)=\\ \mathbb{E}_{y\sim P_Y(y)}\left [(D_Y(y)-\mathbb{E}_{x\sim P_X(x)}D_Y(G_{X\rightarrow Y}(x))-1)^{2}  \right ] + \\ \mathbb{E}_{x\sim P_X(x)}\left [(D_Y(G_{X\rightarrow Y}(x))-\mathbb{E}_{y\sim P_Y(y)}(D_Y(y))+1)^{2}  \right ]
	\end{gathered}	
\end{equation}
\begin{equation}
	\begin{gathered}	
		\mathcal{L}_{Radv}(G_{X\rightarrow Y})= \\ \mathbb{E}_{x\sim P_X(x)}\left [(D_Y(G_{X\rightarrow Y}(x)))-\mathbb{E}_{y\sim P_Y(y)}D_Y(y)-1)^{2}  \right ] +\\ \mathbb{E}_{y\sim P_Y(y)}\left [(D_Y(y)-\mathbb{E}_{x\sim P_X(x)}(D_Y(G_{X\rightarrow Y}(x)))+1)^{2}  \right ]
	\end{gathered}
\end{equation}
where $x$ and $y$ is the magnitude spectrum of noisy speech and that of clean speech (e. g. $\left | {X}_{t,f} \right |$, $\left | {S}_{t,f} \right |$), respectively. Here, $\mathcal{L}_{Radv}(D_Y)$ indicates the adversarial loss of discriminator $D_Y$, and $\mathcal{L}_{Radv}(G_{X\rightarrow Y})$ indicates the adversarial loss of noisy-to-clean generator $G_{X\rightarrow Y}$. In the above equations, the generator $G_{X\rightarrow Y}$ tries to generate the enhanced magnitude spectra that can deceive the discriminator $D_Y$, and $D_Y$ attempts to find the best decision boundary between the clean magnitude spectra $\left | {S}_{t,f} \right |$ and enhanced ones $G_{X\rightarrow Y}(\left | {X}_{t,f} \right |)$. Similarly, we impose two relativistic average adversarial losses $\mathcal{L}_{Radv}(D_X)$ and $\mathcal{L}_{Radv}(F_{Y\rightarrow X})$ for the inverse noisy-to-clean mapping.

\textbf{Cycle-consistency loss:} Due to high randomness, $G$ can map noisy feature space to any random permutation of the clean feature space with only adversarial loss, so any learned mapping functions can produce an output distribution that matches the target distribution. Hence, we apply the cycle-consistency loss to limit space of possible mapping functions and preserve speech context integrity, which can be defined as follows:
\begin{equation}
	\begin{gathered}
		\mathcal{L}_{cycle}(G_{X\rightarrow Y},F_{Y\rightarrow X})=\mathbb{E}_{x\sim P_X(x)}\left [\left \|F_{Y\rightarrow X}(G_{X\rightarrow Y}(x)) -x \right \|_{1}\right ] \\+\mathbb{E}_{y\sim P_Y(y)}\left [\left \|G_{X\rightarrow Y}(F_{Y\rightarrow X}(y)) -y \right \|_{1}\right ]
	\end{gathered}
\end{equation}
where $\left \| \cdot  \right \|_1$ indicates the $L_1$-norm reconstruction error.

\textbf{Identity-mapping loss:} We regularize generators $G$ and $F$ to be close to identity mappings by minimizing identity-mapping loss as in~{\cite{zhu2017unpaired}}, which can be given by:
\begin{equation}
	\begin{gathered}
		\mathcal{L}_{id}(G_{X\rightarrow Y},F_{Y\rightarrow X})=\mathbb{E}_{x\sim P_X(x)}\left [\left \|F_{Y\rightarrow X}(x) -x \right \|_{1}\right ]+ \\ \mathbb{E}_{y\sim P_Y(y)}\left [\left \|G_{X\rightarrow Y}(y) -y \right \|_{1}\right ]
	\end{gathered},
\end{equation}
where magnitude spectrum $y$ and $x$ of the target domain (i.e., $\left | {S}_{t,f} \right |$) and $\left | {X}_{t,f} \right |$) are provided as the input to the generators (i.e., $G_{X\rightarrow Y}$ and $F_{Y\rightarrow X}$), respectively. It helps to preserve the compositions ((i.e., linguistic information) of the source domain and the target domain~{\cite{meng2018cycle}}, enforcing the generators to better map the target distribution simultaneously.

Finally, the total loss function of CycleGAN-MM can be summarized as follows:
\begin{equation}
	\begin{gathered}	
		\mathcal{L}_{CycleGAN-MM}=\mathcal{L}_{Radv}(G_{X\rightarrow Y},D_Y) + \mathcal{L}_{Radv}(F_{Y\rightarrow X},D_X) \\ +\lambda _{cycle}\mathcal{L}_{cycle}(G_{X\rightarrow Y},F_{Y\rightarrow X}) +\lambda _{id}\mathcal{L}_{id}(G_{X\rightarrow Y},F_{Y\rightarrow X}) \label{cyclegan-mm}
	\end{gathered}
\end{equation}
where $\lambda_{cycle}$ and $\lambda_{id}$ are tunable hyper-parameters, which are initialized as 5 and 10, respectively.

\subsection{Deep Complex Denoising Net (DCD Net)}
After the previous stage of enhancement, we compose the estimated magnitude with the noisy phase to get the coarsely enhanced complex spectrogram. In the second stage, the DCD-Net is proposed to further suppress the background noise and simultaneously recover the clean phase spectrum. As shown in Fig.~{\ref{fig:dcd-net}}, DCD Net consists of eight complex encoder/decoder layers, and six complex Temporal-Frequency self-attention (CT-F SA) blocks.
\subsubsection{Complex-valued encoder-decoder\label{Section411}}
The complex encoder/decoder layers are composed of complex-valued 2D convolutions/deconvolution blocks, followed by complex instance normalization (IN) and complex PReLU (CPRelu)~{\cite{trabelsi2017deep}}. The complex IN and CPRelu operate instance normalization and Parametric Relu activation respectively on both real and imaginary values. We design the complex 2D convolution block according to that in DCUNET~{\cite{choi2019phase}} and DCCRN~{\cite{hu2020dccrn}}. The number of the channels for the encoder layers is (32, 32, 64, 64, 128, 128, 256, 256), while the kernel size and strides are set to (3,~5) and (1,~2) along the time and frequency axis, respectively. In practice, complex 2D convolution can be implemented as four traditional real-valued convolutional operations, which can be presented in Fig.~{\ref{fig:complexcnn}}. For complex-valued convolution operation, we define the input complex vector $X=X_r + jX_i$. Meanwhile, the complex-valued convolutional filter $W$ is defined as $W=W_r + jW_i$, where $W_r$ and $W_i$ represent the real and imaginary parts of a complex convolution kernel, respectively. The complex output receives from the complex 2D convolution operation $W\circledast X$, which can be expressed as,
\begin{equation}
	Y_{out}=(W_r \ast X_r -W_i\ast X_i)+j(W_r \ast X_i + W_i\ast X_r),
\end{equation}
where $Y_{out}\in \mathbb{C}$ is the output of the complex-valued 2D convolution.

\subsubsection {Deep complex T-F attention\label{Section412}}

\begin{figure}[t]
	\centering
	\centerline{\includegraphics[width=1\columnwidth]{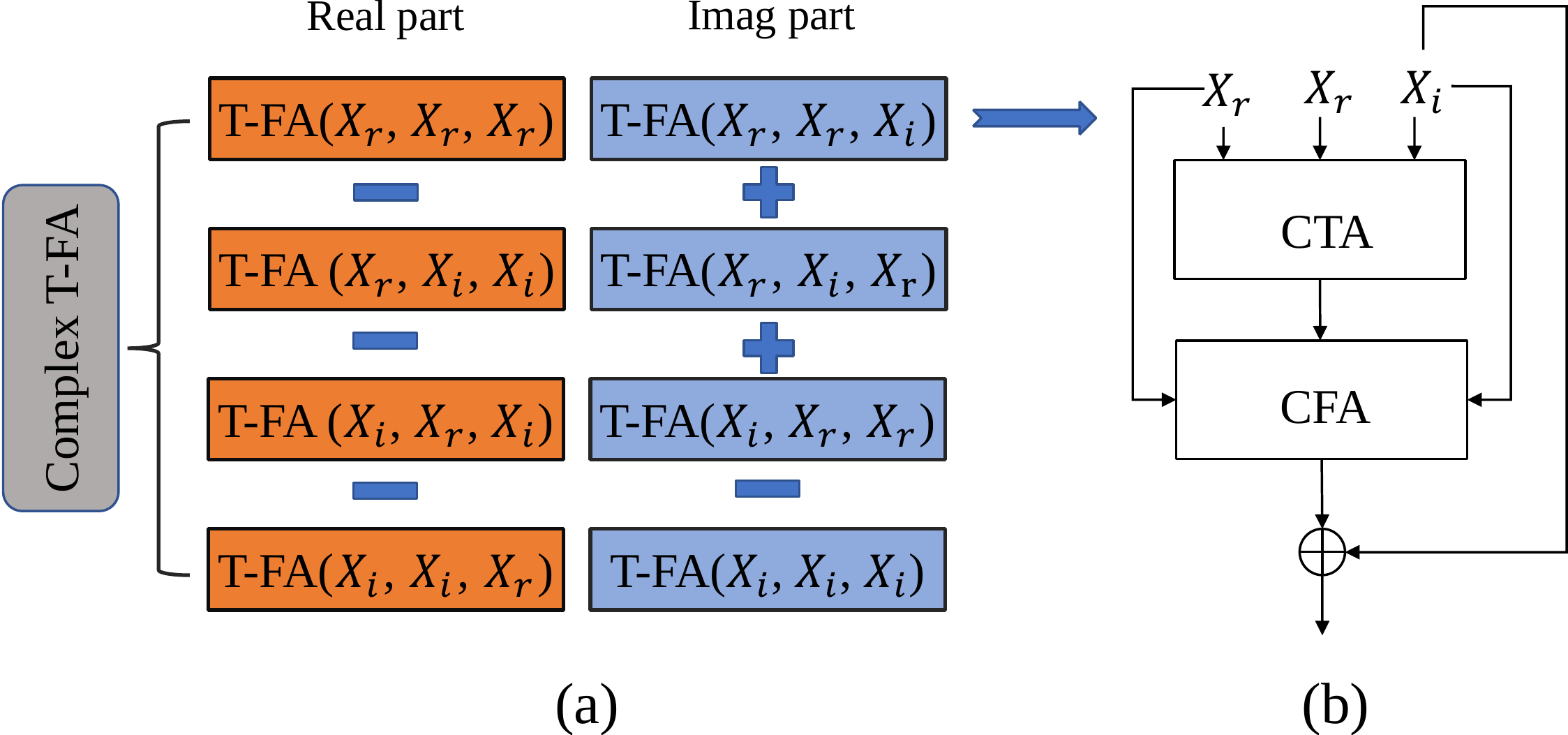}}
	\caption{(a) The calculation of Complex T-F attention. (b) The manner of T-FA($X_r,X_r,X_i$).}
	\label{fig:complexTFA}
	\vspace{-0.2cm}
\end{figure}
\vspace{-0.0cm}
To make DCD Net capable of capturing long temporal dependency with complex-valued features, we introduce complex-valued attention blocks as proposed in~{\cite{yang2020complex}}. Following the temporal-frequency attention as mentioned in Section~{\ref{Section31}}, we propose complex temporal-frequency self-attention (CT-F SA) blocks between the complex encoder-decoder in the DCD-Net, which is calculated by using T-FA in the complex-valued manner. Given the complex-valued input $X=X_r + jX_i$, we first project them to the query matrix $Q_c$, the key matrix $K_c$ and the value matrix $V_c$, and then calculate the complex attention output, which is defined as:
\begin{equation}
	\begin{aligned}	
		Q_c&=W_Q\ast X, K_c=W_K\ast X, V_c=W_V\ast X,\\
		\beta_c&= softmax(Q_c\cdot( K_c)^{T})=\beta_r + j\beta_i\\
		&=softmax((W_Q\ast X_r + jW_Q\ast X_i)\cdot(W_K\ast X_r + jW_K\ast X_i)^T) ,\\
		O_c &= \beta_c\cdot V_c =(\beta_r + j\beta_i) \cdot(W_V\ast X_r + jW_V\ast X_i),\\
		y_c &= \lambda O_C+ K_c = y_r +jy_i,
	\end{aligned}	
\end{equation}
where $W_Q$, $W_K$, and $W_V$ denote the $1\times1$ real-valued convolutional filter. $y_r$ and $y_i$ denote the real and the imaginary part of the complex attention result, respectively. Hence, the complex temporal attention (CTA) can be expressed as:
\begin{equation}
	\begin{aligned}	
		CTA=& (TA(X_r,X_r,X_r)-TA(X_r,X_i,X_i)-TA(X_i,X_r,X_i)\\
           &-TA(X_i,X_i,X_r)) + i(TA(X_r,X_r,X_i)+TA(X_r,X_i,X_r)\\
           &+TA(X_i,X_r,X_r)-TA(X_i,X_i,X_i)),
	\end{aligned}
\end{equation}
where TA denotes the temporal attention as mentioned above. Similarly, the complex frequency attention (CFA) is calculated by using FA in the complex-valued attention manner.

\subsection{Loss function}
The loss function of the proposed two-stage model is defined as below. In the first step, we pretrain the CycleGAN-MM alone with $\mathcal{L}_{CycleGAN-MM}$ until convergence, which is calculated as Eq. ({\ref{cyclegan-mm}}). Then, the CycleGAN-MM and DCD-Net are jointly trained, where the parameters of the first sub-network are initialized with the pretrained CycleGAN-MM, and the overall loss function is expressed as:
\begin{equation}
	\begin{gathered}	
		\mathcal{L}_{DCD}^{Mag}=\left \| \sqrt{\left |\widetilde{S}_r  \right |^2+\left |\widetilde{S}_i  \right |^2 } -  \sqrt{\left |S_r  \right |^2+\left |S_i  \right |^2    } \right \|_2,\\
		\mathcal{L}_{DCD}^{RI}=\left \|\widetilde{S}_r-S_r \right \|_2 +\left \|\widetilde{S}_i-S_i \right \|_2,\\
		\mathcal{L}_{Full}=\mathcal{L}_{DCD}^{RI}+\mathcal{L}_{DCD}^{Mag} + \gamma\mathcal{L}_{CycleGAN-MM},
	\end{gathered}
\end{equation}
where $\mathcal{L}_{DCD}^{Mag}$ and $\mathcal{L}_{DCD}^{RI}$ denote the loss function on the spectral magnitude and RI components, respectively. Here, $\widetilde{S}_r$ and $\widetilde{S}_i$ represent the RI components of the estimated clean speech spectrum, while $S_r$ and $S_i$ represent the target RI components of the clean speech spectrum.

\section{EXPERIMENTS}
\label{Sec4}
\subsection{Datasets\label{Section41}}
In our experiments, we choose two public datasets for comparison. We first evaluated the proposed models as well as several SOTA baselines on a widely used dataset simulated on VoiceBank + DEMAND, and further evaluated our model on the WSJ0-SI84 dataset + DNS challenge.

\textbf{VoiceBank + DEMAND}: This dataset is widely used for evaluation as proposed in~{\cite{valentini2016investigating}}, which is a selection of the Voice Bank corpus with 30 speakers~{\cite{veaux2013voice}}. The training dataset includes 28 speakers’ 11572 utterances in the same accent region (England), while the test set contains two speakers’ (one male and one female) 824 utterances. The total duration of the training set is around 10 hours and the duration of the test set is around 30 mins. For both the training and testing sets, the average speech signal length was three seconds. For the training set, audio samples are added with one of the 10 noise types (2 artificial and 8 from the DEMAND database~{\cite{thiemann2013diverse}}) at four SNRs of 0, 5, 10 and 15 dB. The noise is from different environments including offices, public spaces, transportation stations, and streets. The test set is created with 5 test-noise types (all from the DEMAND database, but totally unseen in the training set) at SNRs of 2.5, 7.5, 12.5 and 17.5 dB. The five types of chosen noise are living room, office, bus, and street noise.

\textbf{WSJ0-SI84 + DNS challenge}: For further evaluation, we use the WSJ0-SI84 dataset~{\cite{paul1992design}}, which includes 7138 utterances by 83 speakers (42 males and 41 females). In our experiments, we split 5428 and 957 utterances by 77 speakers for the training set and validation set, respectively. For the test set, we use two types with 150 utterances of each (seen and unseen speakers). For the first type, the speakers are totally unseen in the training set, while the speakers are within the training dataset for the second type. For the mixed set, we randomly select 20000 noises from the DNS-Challenge~{\footnote{https://github.com/microsoft/DNS-Challenge}} to obtain a 55 hours noise set for training. During each mixed process, a random cut vector of noise is mixed with randomly selected clean utterances. Hence, we established a 15000, 1500 noisy-clean pairs at the SNR range of [5dB,- 4dB, -3dB, -2dB, -1dB, 0dB] for training and validation, respectively. The total duration for the training set is about 30 hours. For testing, we select two noises (i.e., babble and factory1 ) from NOISEX92~{\cite{varga1993assessment}} to obtain totally 900 utterances (450 for seen speakers, 450 for unseen speakers) at three SNRs of -5dB, 0dB and 5dB.

\subsection{Implementation Setup\label{Section42}}
The original raw waveforms were downsampled from 48kHz to 16kHz. The Hanning window of length 20ms is utilized to produce a set of time frames, with 50\% overlap between adjacent frames and the STFT length is 320. When training the VoiceBank dataset, we randomly crop a fixed-length segment (128 frames) with batch size set to 8. As for the WSJ0-SI84 dataset, the maximum utterance length is chunked to 8 seconds and the batch size is set to 16. We adopt the Adam optimizer~{\cite{kingma2014adam}} with the momentum term $\beta_{1}=0.9$, $\beta_{2}=0.999$. In the first stage (20 epochs), we only train the CycleGAN-MM with an initial learning rate(LR) of 0.0002 for discriminators and 0.0005 for generators, respectively. We use $\mathcal{L}_{id}$ only for the first 20 epochs to guide the composition reservation. In the second stage, DCD-Net model is jointly trained with pre-trained CycleGAN-MM, while the learning rates are set to 0.001 and 0.0001 for DCD-Net and CycleGAN-MM, respectively. The same learning rates are maintained for the first 50 epochs, while they linearly decay in the remaining iterations. We train the proposed model for 80 epochs on WSJ0-SI84 + DNS dataset and 100 epochs on Voice Bank + DEMAND dataset, respectively.

\subsection{Evaluation Metrics\label{Section43}}
We use the following metrics to evaluate the objective and subjective quality of the enhanced speech. The objective metrics measure the similarity between the enhanced signal and the clean reference of the test set files. The subjective quality is evaluated by DNSMOS~{\cite{reddy2020dnsmos}}, which is a robust nonintrusive perceptual speech quality metric designed to stack rank noise suppressors with great accuracy. Higher values of all metrics indicate better performance.
\begin{enumerate}[1)]
\item PESQ: Perceptual evaluation of speech quality (PESQ)~{\cite{rix2001perceptual}} score is the most commonly used metric to evaluate speech quality, especially using the wide-band version recommended in ITU-T P.862.2(from -0.5 to 4.5).
\item STOI: Short-Time Objective Intelligibility (STOI)~{\cite{taal2010short}} is used as a robust measurement index for nonlinear processing of noisy speech, e.g., noise reduction on speech intelligibility. The value of STOI ranges from 0 to 1.
\item SSNR: Segmental signal-to-noise ratio, ranging from 0 to $\infty$.
\item CSIG: The mean opinion score (MOS) predicts the speech signal distortion, ranging from 1 to 5~{\cite{hu2007evaluation}}.
\item CBAK: The MOS predicts the intrusiveness of background noise, ranging from 1 to 5~{\cite{hu2007evaluation}}.
\item COVL: The MOS predicts the overall effect, ranging from 1 to 5~{\cite{hu2007evaluation}}.
\item DNSMOS: The speech quality ratings of the processed clips varied from very poor
(MOS=1) to excellent (MOS=5)~{\cite{reddy2020dnsmos}}.
\end{enumerate}

\subsection{Baselines\label{Section44}}
For the comparison on Voice Bank + DEMAND dataset, we adopt both GAN-based and Non-GAN-based methods, which are summarized as follows:
\begin{enumerate}[a)]
\item \textbf{SEGAN}~{\cite{pascual2017segan}} is the first SE approach based on the adversarial framework and works end-to-end with the raw audio. It applies to skip connection to generators, connecting each encoding layer to its homologous decoding layer.
\item \textbf{MMSE-GAN}~{\cite{soni2018time}} introduces a time-frequency masking-based SE approach based on a modified GAN and learns the mask implicitly while predicting the clean T-F representation.
\item \textbf{RSGAN} and \textbf{RSaGAN}~{\cite{baby2019sergan}} introduce relativistic GANs with a relativistic cost function at its discriminators and use gradient penalty to improve speech enhancement performance in the time domain. Note that RSGAN-GP employs relativistic binary cross-entropy loss while RaSGAN-GP employs relativistic average binary cross-entropy loss.
\item \textbf{CP-GAN}~{\cite{liu2020cp}} is a novel GAN-based SE system for coarse-to-fine noise suppression, which contains a densely-connected feature pyramid generator and a dynamic context granularity discriminator.
\item \textbf{MetricGAN}~{\cite{fu2019metricgan}} aims to optimize the generator with respect to one or multiple evaluation metrics such as PESQ and STOI, thus guiding the generators in GANs to generate data with improved metric scores.
\item \textbf{Wave-U-Net}~{\cite{stoller2018wave}} uses the U-Net architecture for speech enhancement, which performs end-to-end audio source separation directly in the time domain.
\item \textbf{LSTM}~{\cite{weninger2014discriminatively}} and \textbf{BiLSTM}~{\cite{erdogan2015phase}} are two RNN-based speech enhancement approaches. Both of them have two layers of RNN cells and the third layer of fully connected NNs.
\item \textbf{DFL-SE}~{\cite{germain2019speech}} proposes a fully-convolutional context aggregation network using a deep feature loss at the raw waveform level, which is based on comparing the internal feature activations in a different network.
\item \textbf{CRN-MSE}~{\cite{tan2018convolutional}} is a typical convolutional recurrent network with encoder-decoder architecture, and MSE loss is used on the estimated and clean log-magnitude spectrogram. Note that we directly use the reported scores of RSGAN, RaSGAN, LSTM, BiLSTM and CRN-MSE from~{\cite{zhang2020loss}}.
\item \textbf{TFSNN}~{\cite{yuan2020time}} proposes a time-frequency smoothing neural network for SE, which effectively models the correlation in the time and frequency dimensions by using LSTM and CNN, respectively.
\end{enumerate}

To further evaluate the proposed method at different SNRs On WSJ0-SI84 + DNS, we re-implement three state-of-the-art baselines, namely \textbf{Noncausal-GCRN}~{\cite{tan2019learning}}, \textbf{DCUNET-20}~{\cite{choi2019phase}} and \textbf{Noncausal-DCCRN}~{\cite{hu2020dccrn}}. Noncausal-GCRN~{\cite{tan2019learning}} is a complex spectral mapping network based on CRN, where both the real and imaginary components are estimated. Notably, to make a fair comparison, we reimplement GCRN as a non-causal configuration, where the Bi-directional LSTM is utilized for time sequencing. DCUNET-20~{\cite{choi2019phase}} adopts the complex-valued building blocks and bounded CRM to deal with the complex-valued spectrum. We use the structure of DCUNET-20 with the stride along with the time dimension set to 1. The channels in encoders are set to [8,16,32,32,64,64,128,128,256,256]. Noncausal-DCCRN~{\cite{hu2020dccrn}} introduces a deep complex convolution recurrent network for speech enhancement, where both CNN and RNN structures handle complex-valued operations. We reimplement DCCRN and employ the Bi-directional LSTM to train the model with a non-causal configuration. Note that we also implement $tanh$ bounded CRM and optimize the system with SI-SNR loss. \\

\renewcommand\arraystretch{1}
\begin{table}[t!]
	\caption{Ablation study on both CycleGAN-MM and CycleGAN-DCD. SA and w/o denote self-attention and without any attention mechanism, respectively. LS and RaLS in stage one denote least-square loss and relativistic average least-square loss, while RI and Mag in stage two denote the MSE loss of RI components and spectral magnitude, respectively..}
	\normalsize
	\centering
	\resizebox{0.9\textwidth}{!}{
		\begin{tabular}{ccccccccc}
			\toprule
			\textbf{Models} & \textbf{Attention}  &\textbf{Loss} & \textbf{PESQ} & \textbf{SSNR} & \textbf{STOI(\%)} & \textbf{CSIG} & \textbf{CBAK} & \textbf{COVL} \\
			\midrule
			Unprocessed &N/A &N/A &1.97 &1.68 &92.1 &3.35 &2.44 &2.63\\
			\midrule
			\multicolumn{6}{c}{\textbf{one-stage System}} \\
			\midrule
			CycleGAN-MM(I) &SA &LS &2.57 &5.69 &92.7 &3.85 &3.11 &3.10 \\
			\midrule
			CycleGAN-MM(II) &SA &RaLS &2.64 &6.18 &92.8 &3.93 &3.17 &3.19 \\
			\midrule
			CycleGAN-MM(III) &T-FA &RaLS &2.69 &6.72 &93.2 &3.98 &3.16 &3.27 \\
			\midrule
			CycleGAN-CM &T-FA &RaLS &2.24 &3.37 &92.0  &3.54 &2.68 &2.71\\
			\midrule
			\multicolumn{6}{c}{\textbf{two-stage System}} \\
			\midrule
			CycleGAN-DCD(I) &w/o &RI &2.72  &8.21 &92.9 &3.91 &3.41 &3.29\\
			\midrule
			CycleGAN-DCD(II) &w/o &RI+Mag & 2.82 &8.92 & 94.1 &4.14 &3.48 &3.39\\
			\midrule
			CycleGAN-DCD(III) &CT-F SA &RI+Mag  &\textbf{2.90} &\textbf{9.33} &\textbf{94.3} &\textbf{4.24} &\textbf{3.57} &\textbf{3.49}\\
			\bottomrule
	\end{tabular}}
	\label{tbl:ablation-study1}
	\vspace{-0.4cm}
\end{table}
\vspace{0cm}

\section{Results and Analysis\label{Section5}}
\label{Sec5}\vspace{-1mm}
\subsection{Ablation study\label{Section51}}
We first investigate the effectiveness of different attention mechanisms and loss functions based on VoiceBank + DEMAND dataset. As shown in Table~{\ref{tbl:complexNN-results}}, we take the CycleGAN with self-attention (SA) and least-square (LS) GAN loss as the baseline. Besides, we implement a CycleGAN-based one-stage network for complex spectral mapping (dubbed CycleGAN-CM), which set the same configuration as CycleGAN-MM(III) except two decoders to separately decode real and imaginary components. From the results, we can have the following observations.
When only the first stage is trained, CycleGAN-MM using RaLS loss achieves 0.07 PESQ, 0.49dB SSNR, 0.08 CSIG, 0.06 CBAK and 0.09 COVL improvements over using conventional LS loss, while using T-FA results in better performance than using SA. CycleGAN-MM(III) integrating RaLS loss and T-FA obtains 0.12 PESQ, 1.03dB SSNR, 0.5\% STOI, 0.13 CSIG, 0.05 CBAK and 0.17 COVL improvements than CycleGAN-MM(I). Additionally, when handling one-stage complex spectral mapping, CycleGAN-CM achieves worse performance than CycleGAN-MM. This reveals the difficulty for conventional GAN-based systems to deal with directly mapping complex spectrum, which is likely caused by the unstable training of the generators and discriminators. In other words, optimizing both RI components is an intractable challenge for the GAN-based SE approaches, and thus achieving worse performance than only estimating the magnitude.\\
Subsequently, in the two-stage systems, DCD-Net is jointly trained with CycleGAN-MM(III) as the first-stage system. When only using RI components loss in DCN-Net, we obtain a marginal improvement over CycleGAN-MM on reducing background noise (e.g., 1.49dB SSNR and 0.25 CBAK improvements), while achieving similar PESQ and STOI. This reveals the necessity and significance of the second stage in residual noise suppression. After adding the magnitude MSE loss of the estimated and clean spectrum, CycleGAN-DCD(II) improves PESQ by 0.10, STOI by 1.2\% and CSIG by 0.23, respectively. This indicates that using both RI and Mag loss in DCD-Net achieves a better performance on speech quality (i.e., PESQ), speech intelligibility(i.e., STOI) and speech distortion(i.e., CSIG). Note that both CycleGAN-DCD(I) and CycleGAN-DCD(II) are trained without any attention block in DCD-Net. Compared with CycleGAN-DCD(II), CycleGAN-DCD(III) trained with CT-F SA blocks provides 0.08 gain on PESQ, 0.41dB gain on SSNR, 0.10 gain on CSIG, 0.09 gain on CBAK and 0.10 gain on COVL, respectively. These results verify the effectiveness of the proposed attention mechanism and loss function for improving speech quality in terms of all objective metrics.

\begin{figure}
	\centering
	\centerline{\includegraphics[width=0.7\columnwidth]{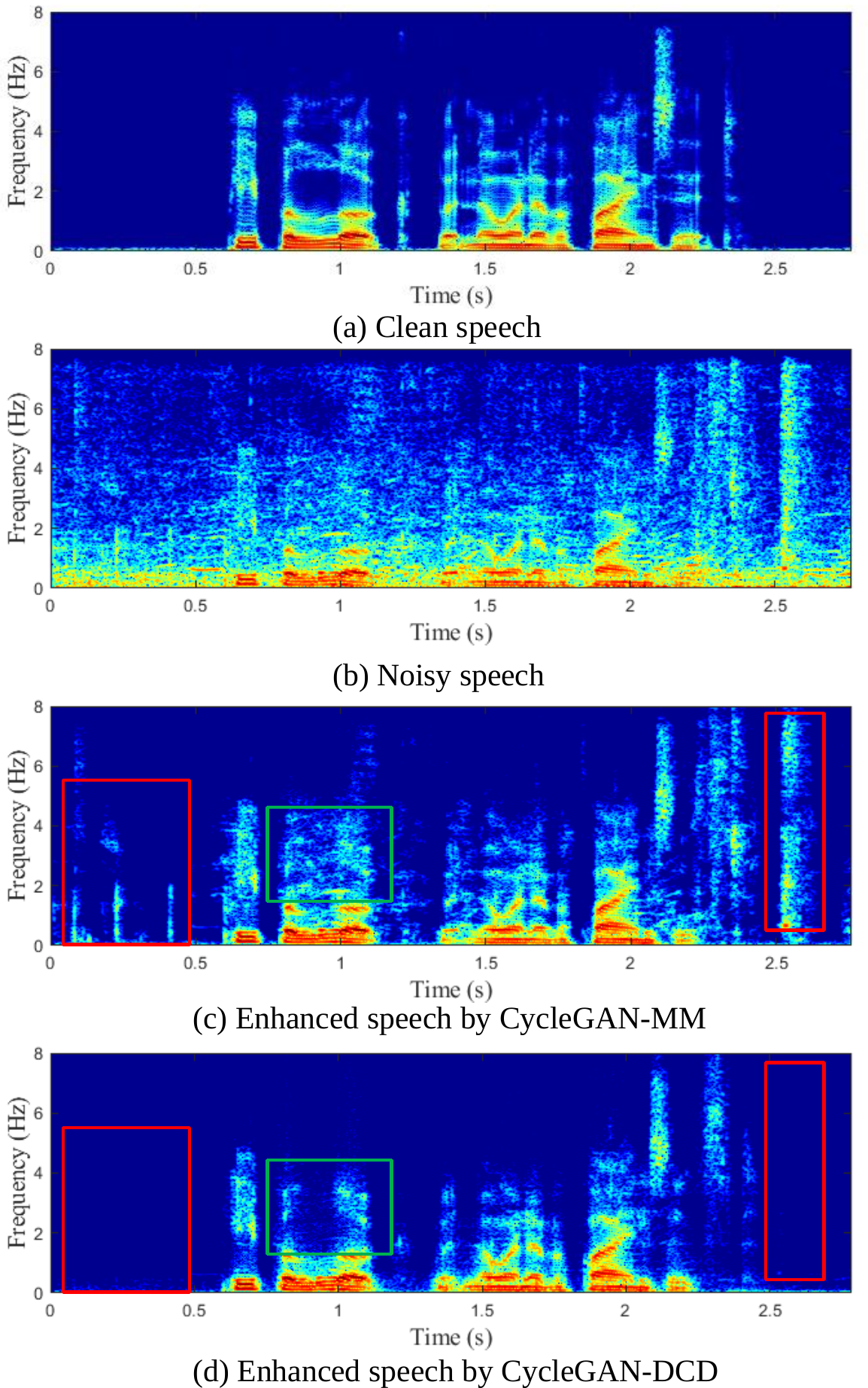}}
	\caption{Illustration of the enhanced results using our proposed models. (a) The speech spectrogram of noisy speech. (b) The speech spectrogram of noisy utterance. (c) The speech spectrogram of the enhanced utterance by CycleGAN-MM. (d) The speech spectrogram of the enhanced utterance by CycleGAN-DCD. }
	\label{fig:p232_010_CycleGAN}
	\vspace{0.5mm}
\end{figure}

Fig.~{\ref{fig:p232_010_CycleGAN}} shows the spectrograms of the clean utterance, the noisy utterance and the utterance enhanced by CycleGAN-MM and CycleGAN-DCD. From the figure, we observe that CycleGAN-DCD can effectively suppress the noise components, which are intractable to be eliminated in CycleGAN-MM. For example, as shown in the red sign area of Fig.~{\ref{fig:p232_010_CycleGAN}} (c) and (d), CycleGAN-DCD achieved better performance under the pure background noise condition. Besides, the green sign area shows CycleGAN-DCD can also effectively suppress the unnatural residual noise while well preserving the speech components in the case of background noise and speech are heavily mixed.

\renewcommand\arraystretch{1}
\begin{table}[t!]
	\caption{Comparison with different complex-spectrum networks and two-stage combinations. }
	\label{tbl:complexNN-results}
	\centering
	\small
	\scalebox{0.8}{
		\begin{tabular}{lcccccc}
			\hline
			\multicolumn{1}{l|}{\textbf{Methods}} & \textbf{PESQ} & \textbf{STOI(\%)} & \textbf{CSIG} & \textbf{CBAK} & \textbf{COVL} \\ \hline
			\multicolumn{1}{l|}{Noisy}  & 1.97 & 92.1 & 3.35 & 2.44 & 2.63 \\ \hline
			\multicolumn{6}{c}{\textbf{One-stage complex Systems}} \\ \hline		
			\multicolumn{1}{l|}{Noncausal-GCRN~{\cite{tan2019learning}}}  & 2.51  & 93.7 & 3.71 & 3.21 & 3.09 \\\hline
			\multicolumn{1}{l|}{Noncausal-DCCRN~{\cite{hu2020dccrn}} } & 2.72 & 93.9 & 3.90 & 3.20 & 3.29 \\\hline
			\multicolumn{1}{l|}{DCD-Net } & 2.70 & 93.8 & 3.92 & 3.24 & 3.21 \\ \hline
			\multicolumn{6}{c}{\textbf{Two-stage complex Systems}} \\ \hline	
			\multicolumn{1}{l|}{CycleGAN-GCRN}  & 2.75 & 93.9 & 3.93 & 3.24 & 3.18 \\ \hline
			\multicolumn{1}{l|}{CycleGAN-DCCRN} & 2.84 & 94.1 & \textbf{4.26} & 3.45 & 3.41 \\ \hline
			\multicolumn{1}{l|}{CycleGAN-DCD} & \textbf{2.90} & \textbf{94.3} & 4.24 & \textbf{3.57} & \textbf{3.49} \\ \hline
		
		\end{tabular}
	}
	\vspace{-5mm}
\end{table}

\renewcommand\arraystretch{1.02}
\begin{table}[t!]
	\caption{ Experimental results among different models including GAN-based systems and Non-GAN-based systems on VoiceBank + DEMAND dataset. We directly use previously reported results. N/A denotes the result is not provided in the original paper. }
	\vspace{5mm}
	\label{tbl:VB-results}
	\centering
	\small
	\scalebox{0.8}{
		\begin{tabular}{lcccccc}
			\hline
			\multicolumn{1}{l|}{\textbf{Methods}} &\textbf{Causality} & \textbf{PESQ} & \textbf{STOI(\%)}  & \textbf{CSIG} & \textbf{CBAK} & \textbf{COVL} \\ \hline
			\multicolumn{1}{l|}{Noisy} & -- & 1.97 & 92.1 & 3.35 & 2.44 & 2.63 \\ \hline
			\multicolumn{6}{c}{\textbf{GAN-based Systems}} \\ \hline
			\multicolumn{1}{l|}{SEGAN~{\cite{pascual2017segan}}} & $\times$ & 2.16 & 92.5 & 3.48 & 2.94 & 2.80 \\ \hline
			\multicolumn{1}{l|}{MMSEGAN~{\cite{soni2018time}}} & $\times$ & 2.53 & 93.0 & 3.80 & 3.12 & 3.14 \\ \hline
			\multicolumn{1}{l|}{RSGAN~{\cite{baby2019sergan}}}  & $\times$ & 2.51 & 93.7 & 3.78 & 3.23 & 3.16 \\ \hline
			\multicolumn{1}{l|}{RaSGAN ~{\cite{baby2019sergan}}} &  $\times$ & 2.57 & 93.7 & 3.83 & 3.28 & 3.20 \\ \hline
			\multicolumn{1}{l|}{CP-GAN~{\cite{liu2020cp}} } &  $\times$ & 2.64 & 94.0 & 3.93 & 3.29 & 3.28 \\ \hline
			
			\multicolumn{1}{l|}{MetricGAN~{\cite{fu2019metricgan}}} &$\times$ & 2.86 & N/A & 3.99 & 3.18 & 3.42 \\ \hline
			\multicolumn{6}{c}{\textbf{Non-GAN-based Systems}} \\ \hline			
			\multicolumn{1}{l|}{Wave-U-Net~{\cite{stoller2018wave}}} & $\times$ & 2.64 & N/A & 3.56 & 3.08 & 3.09 \\  \hline
			\multicolumn{1}{l|}{LSTM ~{\cite{weninger2014discriminatively}}} & $\checkmark$ & 2.56 & 91.4 & 3.87 & 2.87 & 3.20 \\ \hline
			\multicolumn{1}{l|}{BiLSTM~{\cite{erdogan2015phase}}} &  $\times$ & 2.70 & 92.5 & 3.99 & 2.95 & 3.34 \\ \hline
			\multicolumn{1}{l|}{DFL-SE~{\cite{germain2019speech}} } &  $\checkmark$ &N/A & N/A & 3.86 & 3.33 & 3.22 \\ \hline
			\multicolumn{1}{l|}{CRN-MSE~{\cite{tan2018convolutional}} } & $\checkmark$ & 2.74 & 93.4 & 3.86 & 3.14 & 3.30 \\ \hline
			\multicolumn{1}{l|}{TFSNN~{\cite{yuan2020time}} } & $\checkmark$ & 2.79 & N/A & 4.17 & 3.27 & \textbf{3.49} \\ \hline
			
			\multicolumn{6}{c}{\textbf{Proposed CycleGAN-based approaches}} \\ \hline
			\multicolumn{1}{l|}{CycleGAN-MM} &  $\times$  & 2.69 & 93.2 & 3.98 & 3.16 & 3.27 \\ \hline
			\multicolumn{1}{l|}{CycleGAN-DCD} &  $\times$ & \textbf{2.90} & \textbf{94.3} & \textbf{4.24} & \textbf{3.57} & \textbf{3.49} \\ \hline
		\end{tabular}
	}
	\vspace{-5mm}
\end{table}

\subsection{Comparison with different two-stage structures\label{Section53}}

To validate the efficacy of DCD-Net and the proposed two-stage complex structure, we also conduct experiments on VoiceBank + DEMAND
dataset. Specifically, we first compare DCD-Net with other existing complex spectrum enhancing networks (i.e., Noncausal-GCRN~{\cite{tan2019learning}}, Noncausal-DCCRN~{\cite{hu2020dccrn}}) and then design their corresponding two-stage structures (i.e., CycleGAN-GCRN, CycleGAN-DCCRN) for comparison. Note that the trainable parameter of Noncausal-GCRN, Noncausal-DCCRN and DCD-Net is 9.8 million, 3.7 million and 3.5 million, respectively. As shown in Table~{\ref{tbl:objective-results}}, one can observe following phenomena. Firstly, when compared with real-valued complex-mapping network, DCD-Net outperforms GCRN in all metrics by a large margin. For example, DCD-Net provides average 0.19 PESQ, 0.21 CSIG, 0.03 CBAK and 0.12 COVL improvements than GCRN with relatively lower model complexity, which indicates the merit of complex-valued networks. Secondly, DCD-Net surpasses another complex-valued network (i.e., DCCRN) in terms of COVL and CBAK scores, while providing similar PESQ and STOI scores. This indicates DCD-Net can effectively suppress the background noise and reduce the speech distortion simultaneously. Finally, to demonstrate the merit of the proposed two-stage combination, we combine CycleGAN-MM with GCRN, DCCRN and DCD-Net as different two-stage structures (i.e., CycleGAN-GCRN, CycleGAN-DCCRN and the proposed CycleGAN-DCD) for comparison. Note that we also use bounded CRM in CycleGAN-GCRN for better performance. Compared with different two-stage methods, CycleGAN-DCD achieves consistently better performances than real-valued CycleGAN-GCRN by a significant margin, while CycleGAN-DCD outperforms complex-valued CycleGAN-DCCRN in terms of PESQ, STOI, CBAK and COVL. This validates that the proposed CycleGAN-DCD surpasses other two-stage structures including real-valued and complex-valued methods.

\subsection{Comparison with the State-of-the-Art\label{Section54}}

\renewcommand\arraystretch{1.62}
\begin{sidewaystable}[htpb]
	\caption{Objective result comparisons among different models in terms of PESQ, STOI and SSNR for both seen and unseen speaker cases. \textbf{BOLD} indicats the best score in each case.}
	\Huge
	\centering
	\label{tbl:objective-results}
	\resizebox{1.05\textwidth}{!}{
		\begin{tabular}{cc|c|cccccccc|cccccccc|cccccccc}
			\hline
			&\textbf{Metrics}
			&\multirow{3}*{\rotatebox{90}{\textbf{Causality}}}
			&\multicolumn{8}{c|}{\textbf{PESQ}} &\multicolumn{8}{c|}{\textbf{STOI(\%)}} &\multicolumn{8}{c}{\textbf{SSNR(dB)}} \\
			\cline{1-2}\cline{4-27}
			& \textbf{Speakers} & &\multicolumn{4}{c|}{\textbf{Seen}} &\multicolumn{4}{c|}{\textbf{Unseen}} &\multicolumn{4}{c|}{\textbf{Seen}} &\multicolumn{4}{c|}{\textbf{Unseen}} &\multicolumn{4}{c|}{\textbf{Seen}} &\multicolumn{4}{c}{\textbf{Unseen}} \\
			\cline{1-2}\cline{4-27}
			& \textbf{SNR(dB)}  & &-5 &0 &5 &\multicolumn{1}{c|}{Avg.} &-5 &0 &5 &Avg. &-5 &0 &5 &\multicolumn{1}{c|}{Avg.} &-5 &0 &5 &\multicolumn{1}{c|}{Avg.} &-5 &0 &5 &\multicolumn{1}{c|}{Avg.} &-5 &0 &5 &Avg.\\
			\cline{1-27}
			\multirow{6}*{\rotatebox{90}{\textbf{Babble}}}
			&\multicolumn{1}{|c|}{\textbf{Noisy}} &N/A &1.41 &1.56 &1.80 &\multicolumn{1}{c|}{1.60} &1.38 &1.52 &1.78 &1.56 &59.10 &74.18 &86.38 &\multicolumn{1}{c|}{73.49} &60.25 &73.65 &85.30 &73.07 &-6.70 &-4.09 &-0.97 &\multicolumn{1}{c|}{-3.92} &-6.77 &-4.21 &-1.12 &-4.03\\
			\cline{2-27}
			&\multicolumn{1}{|c|}{\textbf{DCUNET}~{\cite{choi2019phase}}}  &$\times$ &1.75 &2.34 &2.88 &\multicolumn{1}{c|}{2.32} &1.71 &2.23 &2.82  &2.25 &79.28 &91.82 &95.74 &\multicolumn{1}{c|}{88.95} &78.53 &91.26 &95.28  &88.36 &1.98 &4.12 &5.89  &\multicolumn{1}{c|}{3.95} &2.01 &4.06 &5.79 &4.01\\
			\cline{2-27}
			&\multicolumn{1}{|c|}{\textbf{Noncausal-DCCRN}~{\cite{hu2020dccrn}}}  &$\times$ &1.77 &2.39 &2.91 &\multicolumn{1}{c|}{2.36} &1.73 &2.26 &2.83  &2.27 &82.12 &91.19 &95.97 &\multicolumn{1}{c|}{89.76} &81.72 &91.13 &95.93 &89.59 &1.19 &2.47 &4.40  &\multicolumn{1}{c|}{2.67} &1.32 &2.38 &4.33 &2.69\\
			\cline{2-27}
			&\multicolumn{1}{|c|}{\textbf{Noncausal-GCRN}~{\cite{tan2019learning}}} &$\times$ &1.83 &2.29 &2.88 &\multicolumn{1}{c|}{2.34} &1.78 &2.29 &2.87 &2.31 &83.09 &92.55 &96.32 &\multicolumn{1}{c|}{90.65} &82.61 &92.51 &96.22 &90.44 &2.15 &4.47 &6.95 &\multicolumn{1}{c|}{4.53} &2.16 &4.84 &6.97 &4.66\\
			\cline{2-27}
			&\multicolumn{1}{|c|}{\textbf{CycleGAN-ME(Pro.)}} &$\times$ &1.55 &2.04 &2.60 &\multicolumn{1}{c|}{2.06} &1.52 &2.02 &2.58 &2.04 &76.02 &89.54 &94.98 &\multicolumn{1}{c|}{86.85} &75.70 &89.72 &95.03 &86.82 &1.12 &3.82 &6.22 &\multicolumn{1}{c|}{3.72} &1.02 &3.84 &6.34 &3.74\\
			\cline{2-27}
			&\multicolumn{1}{|c|}{\textbf{CycleGAN-DCD(Pro.)}}  &$\times$ &\textbf{1.85} &\textbf{2.37} &\textbf{2.92} &\multicolumn{1}{c|}{\textbf{2.38}} &\textbf{1.81} &\textbf{2.36} &\textbf{2.90} &\textbf{2.35} &\textbf{83.43} &\textbf{92.78} &\textbf{96.34} &\multicolumn{1}{c|}{\textbf{90.86}} &\textbf{82.95} &\textbf{92.73} &\textbf{96.25} &\textbf{90.63} &\textbf{2.30} &\textbf{4.91} &\textbf{7.18} &\multicolumn{1}{c|}{\textbf{4.80}} &\textbf{2.35} &\textbf{5.03} &\textbf{7.23} &\textbf{4.87}\\
			\hline
			
			\multirow{6}*{\rotatebox{90}{\textbf{Factory1}}}
			&\multicolumn{1}{|c|}{\textbf{Noisy}} &N/A &1.35 &1.56 &1.75 &\multicolumn{1}{c|}{1.59} &1.32 &1.47 &1.73 &1.51 &59.93 &74.22 &87.14 &\multicolumn{1}{c|}{73.52} &59.97 &74.22 &86.37 &73.52 &-6.70 &-0.05 &-1.08 &\multicolumn{1}{c|}{0.02} &-6.82 &-4.27 &5.03 &-1.20\\
			\cline{2-27}
			&\multicolumn{1}{|c|}{\textbf{DCUNET}~{\cite{choi2019phase}}} & $\times$ &1.88 &2.36 &2.90  &\multicolumn{1}{c|}{2.37}&1.82 &2.28 &2.85 &2.31 &81.33 &91.60 &95.12 &\multicolumn{1}{c|}{89.25} &80.24 &91.24 &95.58 &89.02 &2.12 &4.60 &6.01 &\multicolumn{1}{c|}{4.23} &2.16 &4.67 &6.12 &4.32\\
			\cline{2-27}
			
			&\multicolumn{1}{|c|}{\textbf{Noncausal-DCCRN}~{\cite{hu2020dccrn}}} & $\times$ &1.89 &2.39 &2.90 &\multicolumn{1}{c|}{2.40} &1.81 &2.30 &2.88  &2.33 &82.21 &91.20 &95.84 &\multicolumn{1}{c|}{89.75} &81.61 &90.88 &96.01 &89.50 &1.42 &2.50 &4.64 &\multicolumn{1}{c|}{2.86} &1.45 &2.72 &4.70 &2.96\\
			\cline{2-27}
			&\multicolumn{1}{|c|}{\textbf{Noncausal-GCRN}~{\cite{tan2019learning}}} & $\times$ &1.90 &2.36 &2.91 &\multicolumn{1}{c|}{2.39} &\textbf{1.92} &2.41 &2.93 &2.42 &82.48 &91.74 &95.92 &\multicolumn{1}{c|}{90.04} &82.44 &92.04 &96.07 &90.18 &2.41 &4.78 &6.70 &\multicolumn{1}{c|}{4.63} &2.94 &4.85 &6.90 &4.90\\
			\cline{2-27}
			
			&\multicolumn{1}{|c|}{\textbf{CycleGAN-MM(Pro.)}} & $\times$ &1.69 &2.16 &2.68 &\multicolumn{1}{c|}{2.18} &1.66 &2.15 &2.65 &2.15 &79.26 &90.37 &96.10 &\multicolumn{1}{c|}{88.58} &78.23 &90.34 &95.20 &87.92 &1.79 &4.27 &6.38 &\multicolumn{1}{c|}{4.14} &1.80 &4.23 &6.45 &4.16\\
			\cline{2-27}
			&\multicolumn{1}{|c|}{\textbf{CycleGAN-DCD(Pro.)}} & $\times$ &\textbf{1.93} &\textbf{2.56} &\textbf{2.96} &\multicolumn{1}{c|}{\textbf{2.49}} &1.91 &\textbf{2.44} &\textbf{2.94} &\textbf{2.44} &\textbf{83.99} &\textbf{92.38} &\textbf{96.25} &\multicolumn{1}{c|}{\textbf{90.88}} &\textbf{82.53} &\textbf{92.09} &\textbf{96.10} &\textbf{90.24} &\textbf{2.51} &\textbf{4.97} &\textbf{6.98} &\multicolumn{1}{c|}{\textbf{4.82}} &\textbf{3.05} &\textbf{5.04} &\textbf{7.18} &\textbf{5.09}\\
			\hline
	\end{tabular}}
	\vspace{-0.4cm}
\end{sidewaystable}

Table~{\ref{tbl:VB-results}} shows the comparisons with mentioned baselines on VoiceBank + DEMAND dataset. Note that CycleGAN-MM and CycleGAN-DCD employ the best configurations from the ablation study. First, we observe that CycleGAN-DCD achieves a notable improvement over most existed GAN-based methods. For example, CycleGAN-DCD exceeds SEGAN by a large margin in PESQ, STOI, CSIG, CBAK and COVL, which are 0.73, 1.8\%, 0.76, 0.63 and 0.69, respectively. Compared with more recently proposed CP-GAN and Metric-GAN, our model still achieves better performance in speech quality and speech intelligibility. Note that the consistent improvements in CSIG, CBAK and COVL also indicate that CycleGAN-DCD performs better in preserving speech integrity while removing the background noise. Then, when it comes to recently proposed Non-GAN methods, the proposed model also achieves better performance across most metrics. For example, CycleGAN-DCD provides 0.38 CSIG, 0.24 CBAK and 0.27 COVL improvements than DFL-SE, while CycleGAN-MM gets lower CBAK and similar COVL over DFL-SE. This indicates the two-stage denoising system demonstrates consistently superior performance on background noise suppression, while further improving the speech distortion and overall effect. \\
Table~{\ref{tbl:objective-results}} shows the comparisons with DCUNET, GCRN and DCCRN on WSJ0-SI84 + DNS dataset. Firstly, we observe that DCUNET, Noncausal-DCCRN and Noncausal-GCRN obtain better performance than CycleGAN-MM under different noise conditions. This is because CycleGAN-MM only estimates the magnitude spectrum and reuses the noisy phase to reconstruct waveform, which causes severe phase distortion under low SNRs. For example, Noncausal-GCRN provides average 0.28 and 0.27 PESQ improvements than CycleGAN-MM on Babble and Factory1 noises, while 3.71\% and 1.86\% improvements in terms of STOI. Secondly, when adding the denoising net to refine the coarsely enhanced complex spectrum, CycleGAN-DCD outperforms the one-stage model by a large margin in all metrics. For example, CycleGAN-DCD provides average 0.32 and 0.30 PESQ improvements than CycleGAN-MM on Babble and Factory1 noises, while providing 0.99dB and 1.03dB gain SSNR. This indicates the necessity and significance of the proposed DCD-Net in improving the speech quality and intelligibility, while further suppressing the residual noise. It can also be observed that the proposed two-stage model consistently outperforms the baselines in terms of all metrics. For example, compared with the best baseline Noncausal-GCRN, we notice that CycleGAN-DCD obtains average 0.06, 0.33\% and 0.21dB improvements in terms of PESQ, STOI and SSNR, respectively.
\begin{figure}[t]
	\centering
	\centerline{\includegraphics[width=80mm]{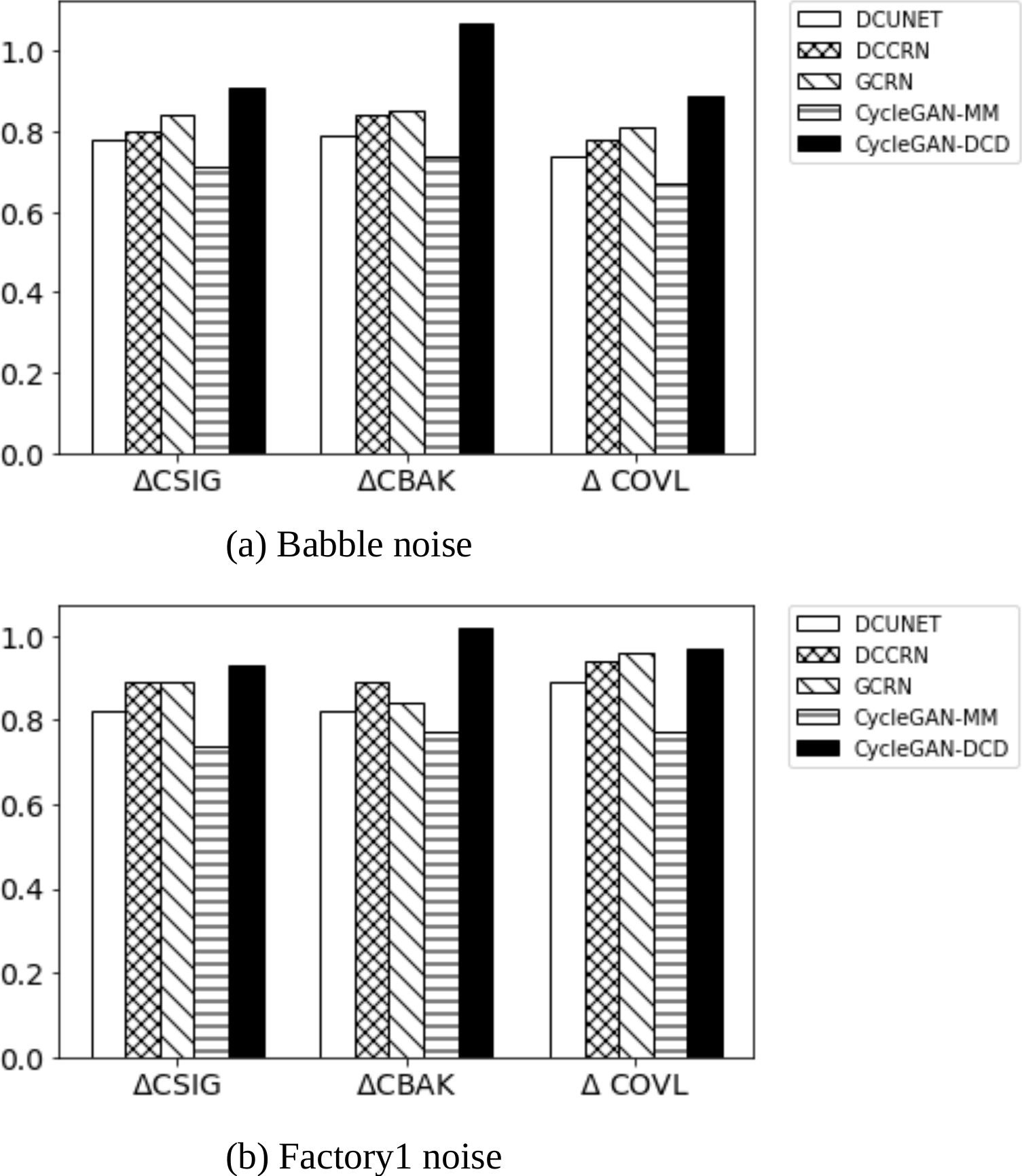}}
	\caption{Average CSIG, CBAK and COVL improvements over the unprocessed mixtures for different noise types at SNRs of -5, 0 and 5dB.}
	\label{fig:CSIG}
	\vspace{-0.4cm}
\end{figure}
\vspace{-0.2cm}

\renewcommand\arraystretch{1.02}
\begin{table}
	\centering
	\caption{The average DNSMOS scores of CycleGAN-DCD and the baseline approaches on the test set at SNRs of -5, 0 and 5dB.} %
	\label{tbl:dnsmos_scores}
	\scriptsize
	\renewcommand\arraystretch{1.2}
	\setlength{\tabcolsep}{6mm}{
		\begin{tabular}{lcccc}
			\toprule
			Noise Type      &\multicolumn{2}{c}{Factory1}  &\multicolumn{2}{c}{Babble}           \\
			\midrule
			Speakers     & seen & unseen & seen & unseen \\
			\midrule
			Mixture     & 2.30          & 2.31     & 2.62   & 2.65         \\
			DCUNET      & 3.12          & 3.13     & 3.16  & 3.16    \\
			Noncausal-DCCRN       &3.15   &3.15  & 3.18   & 3.18           \\
			Noncausal-GCRN        & \textbf{3.30}  & 3.31 & 3.36   & 3.38          \\
			CycleGAN-MM(Pro.)  &3.03   &3.06  & 3.11   & 3.15       \\
			CycleGAN-DCD(Pro.) & \textbf{3.30}  & \textbf{3.33} & \textbf{3.42} & \textbf{3.43} \\
			\bottomrule
		\end{tabular}
	}
\end{table}

Besides, CSIG, CBAK and COVL improvements (i.e., $\Delta$ CSIG, $\Delta$ CBAK and $\Delta$ COVL) over the unprocessed mixtures are shown in Fig.~{\ref{fig:CSIG}}. we can observe that our proposed approach produces considerable improvements than all the baselines in all metrics, especially in CBAK. This reveals the superior capability of CycleGAN-DCD on reducing residual noise and speech distortion, while consistently improving the speech overall quality.

The evaluation of the subjective speech quality on WSJ0-SI84 + DNS dataset is presented in Table~{\ref{tbl:dnsmos_scores}}. We can observe that our method yields the best performance on both seen and unseen speakers with Babble and Factory1 noise types. For example, CycleGAN-DCD yieds average 0.27 and 0.30 DNSMOS scores over one-stage CycleGAN-MM for Factory1 and Babble noise types, respectively. It indicates that our two-stage system can dramatically improve the speech perception of enhanced speech over CycleGAN-MM and other baseline systems under various noisy conditions.

\section{Conclusion and Future work}
\label{Sec6}
In this work, a deep complex-valued denoising sub-net is integrated into a CycleGAN-based magnitude mapping sub-net as a two-stage SE approach, which aims at estimating both the magnitude and phase of the clean speech spectrum. In the first stage, a CycleGAN-based network is first trained to estimate the spectral magnitude with relativistic average least-square losses, cycle-consistency losses and an identity mapping loss. Then, the coarsely estimated magnitude is coupled with the original noisy phase as the input to a complex denoising net, which aims to suppress the residual noise and recovery the clean phase. Notably, the denoising net directly estimates both RI components of the clean spectrum applied with a complex ration mask. Additionally, the temporal-frequency attention mechanism is employed in both two stages for modeling the global dependencies along temporal and frequency dimensions, respectively. To the best of our knowledge, this is the first CycleGAN-based approach to estimate both the clean magnitude and phase information for single-channel SE. Experiments results on VoiceBank and WSJ0-SI84 datasets verify that the proposed method outperforms the conventional one-stage CycleGAN-based SE model and other state-of-the-art GAN-based as well as Non-GAN-based baselines by a considerable margin. \\
In future work, we will investigate the proposed CycleGAN-DCD as a complex spectral mapping network for multi-microphone speech enhancement, in which accurate phase estimation is likely more essential. Considering the promising performance of power compression and phase estimation on speech dereverberation task as discussed in~{\cite{li2021importance}}, we will investigate to use the compressed spectral magnitude as the input feature to the first stage. Besides, we will attempt to decompose the two-stage SE task into two much easier sub-tasks. In the first task, we plan to employ a CycleGAN-based network to transform the non-stationary noise type to stationary noise type like noise-whitening, while we plan to utilize a denoising net to suppress the stationary noise in the second sub-task.





\section*{Acknowledgment}
This work was supported in part by the National Natural Science Foundation of China under Grant 61631016 and Grant 61501410, and in part by the Fundamental Research Funds for the Central Universities under Grant 3132018XNG1805. This work was also supported by the Open Research Project of the State Key Laboratory of Media Convergence and Communication, Communication University of China, China (No. SKLMCC2020KF005)
\bibliography{myrefs}

\end{document}